\documentclass{aa}
\usepackage{graphicx,amsmath}
\usepackage{txfonts}

\usepackage{natbib} 
\bibpunct{(}{)}{;}{a}{}{,}
\begin{document}
\title{Linear wavelength correlation matrices of photospheric and chromospheric spectral lines: 1. Observations vs.~modeling}

   \author{C.Beck\inst{1} \and W.Rammacher\inst{2}}
        
   \titlerunning{Correlation matrices of spectral lines}
  \authorrunning{C.Beck \& W.Rammacher}  
\offprints{C.Beck}

   \institute{Instituto de Astrof\'{\i}sica de Canarias
     (CSIC), La Laguna, Tenerife, Spain\\
     \and Kiepenheuer-Institut f\"ur Sonnenphysik,
      Freiburg, Germany.
     } 
\date{Received xxx; accepted xxx}
\abstract{The process that heats the solar chromosphere is a difficult target
  for observational studies because the assumption of local thermal
  equilibrium (LTE) is not valid in the upper solar atmosphere, { complicating\rm} the analysis of spectra.}{We investigate the linear correlation
  coefficient between the intensities at different wavelengths in photospheric
  and chromospheric spectral lines because the correlation can be determined
  { directly for any spectra from observations or modeling\rm}. Waves that propagate vertically
  through the stratified solar atmosphere affect different wavelengths at
  different times when the contribution functions for each wavelength peak in
  different layers. This leads to a characteristic pattern of (non-)coherence
  of { the intensity at various\rm} wavelengths with respect to each other
  that carries information on the physical processes.}{We derived the 
  correlation matrices for several photospheric (line blends of
  \ion{Ca}{II} H near 396\thinspace nm, 630\thinspace nm, 1082.7\thinspace nm,
  1565\thinspace nm) and chromospheric spectral lines (\ion{Ca}{II} H 396.85\thinspace
  nm, \ion{Ca}{II} IR 854\thinspace nm, \ion{Ca}{II} IR 866\thinspace nm,
  \ion{He}{I} 1083\thinspace nm) from observations. We separated locations
  with significant photospheric polarization signal, and thus, magnetic
  fields, from those { without\rm} polarization signal. For comparison with
  the observations, we calculate correlation matrices for spectra from 
  simplified LTE modeling { approaches\rm}, 1-D NLTE simulations, and a 3-D MHD simulation
  run. We apply the correlation method also to temperature maps at different
  optical depth layers derived from a LTE inversion of \ion{Ca}{II} H
  spectra.}{We find that all photospheric spectral lines show a similar
  pattern: a pronounced asymmetry of the correlation between line core and
  red or blue wing. The pattern { cannot be reproduced with a simulation of the granulation pattern, but with} waves that travel upwards through the formation heights of the lines. { The correct asymmetry between red and blue wing only appears when\rm } a temperature enhancement occurs simultaneously with a downflow velocity { in the wave simulation\rm}. All chromospheric spectral lines show a more complex pattern. In the case of \ion{Ca}{II} H, the 1-D NLTE simulations of monochromatic waves produce a correlation matrix that qualitatively matches to the observations. The photospheric signature is well reproduced in the matrix { derived from the\rm} 3-D MHD simulation.}{The correlation matrices of observed photospheric and chromospheric spectral lines are highly structured with characteristic and different patterns in every spectral line. The comparison with matrices derived from simulations and simple modeling suggests that the main driver of the detected patterns are upwards propagating waves. Application of the correlation method to 3-D temperature cubes seems to be a promising tool for a detailed comparison of simulation results and observations in future studies.}

\maketitle
\begin{figure}
\centerline{\resizebox{7.cm}{!}{\includegraphics{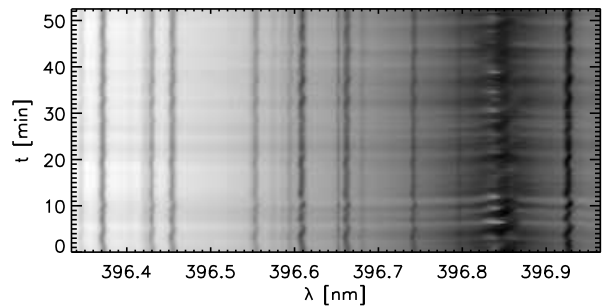}}}$ $\\
\caption{Example of observed \ion{Ca}{II} H spectra. \label{fig1}}
\end{figure}

\section{Introduction}
The \ion{Ca}{II} H and K spectral lines have been one of the most important
chromospheric diagnostics in solar and stellar physics. The lines show a
generally complex behavior with some shared characteristics. Both H and K
exhibit sudden ``bright grains'' (BGs) where wavelengths in and near the line
core revert to emission for a short duration
\citep[e.g.,][]{rutten+uitenbroek1991}. The BGs are often repetitive with a
cadence around 180-200 sec \citep[e.g.,][and references
therein]{beck+etal2008}. As explanation for the BGs the (upwards) propagation
of acoustic waves has been suggested soon after their first detection, because
the BGs can often been seen appearing in the line wing already around 50-100
sec before the emission in the core \citep[see Fig.~\ref{fig1},
or][]{liu1974,beck+etal2008}{, which} supports the idea of a wave propagation.

Due to their formation above the photosphere in a layer with mostly NLTE
conditions, the theoretical treatment and the interpretation of the Ca line
behavior in observations is difficult. { The assumption of
LTE breaks down for the formation of the line cores of the \ion{Ca}{II}
resonance lines, because of the low gas density, but is still valid in the
wings of these lines.}%The assumption of LTE for the atmosphere breaks down due to the low gas density, whereas it still applies partly in the line wing. 

For the analysis of the spectra fortunately also diagnostic methods exist that are independent of the LTE or non-LTE assumption. One of these methods is the matrix of the linear correlation coefficient of the intensity at different wavelengths. The correlation shows a characteristic pattern if there is a causal relationship between different atmosphere layers. When a time lag is introduced in the analysis of a temporal sequence of spectra, the pattern changes accordingly, as an intensity variation at a wavelength $\lambda_1$ at a time $t_1$ appears at a different time $t_2$ in $\lambda_2$. One of the great advantages of the correlation coefficient is that it can be calculated for observed or synthetic spectra from either numerical or analytical models in the same way, for any kind of spectral line or also continuum levels. Interestingly, the correlation matrices can also be derived for other physical quantities, like temperature cuts at geometrical heights in simulations, which widens the field of possible applications of the analysis method.

{ Since} there is no current literature on the theoretical expectations or
the interpretation of { the} correlation matrices { (but see for
example \citet{cram1978} for a similar type of study in the Fourier domain)}, we want only to present
in this contribution some examples of correlation matrices for observations
and { four} numerical experiments with { a} different degree of sophistication. The
equation for calculating the correlation matrices is explained in
Sect.~\ref{secttheo}. We show the correlation matrices for observations of
spectral lines from near-UV (396 nm) to near-IR (1.5 $\mu$m), covering several
photospheric and chromospheric spectral lines, in Sect.~\ref{obsmat}. The
corresponding results for the numerical experiments are shown in
Sect.~\ref{theomat}. Section \ref{secttemp} shows correlation matrices for
temperature as an example of another physical quantity to which the method can
be applied. We summarize and discuss the findings in Sect.~\ref{finalsect},
whereas Sect.~\ref{conclusions} gives any conclusions that can be drawn at this
time.
\section{Calculation of correlation matrices\label{secttheo}}
We use the standard definition for the linear correlation coefficient,
$r_{AB}$ for two { quantities} $A$ and $B$:
\begin{equation}
r_{AB} = \frac{1}{\sqrt{ \int_V A^2\cdot \int_V B^2 } }\int_V (A-\langle A\rangle)\cdot (B-\langle B\rangle) \;, \label{eq1}
\end{equation}
where the integration $\int_V$ is to be executed over all elements of $A$ and $B$, respectively. $\langle \rangle$ denotes the average. %The structures $A$ and $B$ can have any dimension; i.e., they can correspond to 2-D images or 1-D vectors.

For our purpose of wavelength correlations, we normally used a monochromatic
image in one wavelength, $I(\lambda_0)$, as 2-D image $A$, to be correlated
with $B = I(\lambda_1)$ to obtain $r_{\lambda_0\lambda_1}$. We also selected
in some cases subfields of observations or simulations to investigate the
influence of magnetic fields on the correlation matrices. 
%In this case, we collected spectra from the subfield in new 2-D storage
%arrays that did not retain the previous spatial locations of the spectra. For
%the correlation of wavelengths this re-ordering of spectra has no
%consequences and Eq.~(\ref{eq1}) can be applied as well. 
The correlation matrices relating spectra inside the same spectral
range are { square and} symmetric to the diagonal; i.e.,
$r_{\lambda_0\lambda_1} = r_{\lambda_1\lambda_0}$ { (see for example Fig.~\ref{fig6})}. For cross-correlations of different spectral ranges,
{ the resulting matrix is in general rectangular without a symmetry
  axis}. Cross-correlation matrices between different spectral ranges are
especially interesting where time-series of spectra are available, because
they allow to study the variation of the correlation for different time
lags. For propagating waves and a finite difference in formation height, the
maximum correlation should be reached for a time lag corresponding to the wave
travel time between the two height layers. Cross-correlation and correlations with a time lag will be investigated in a subsequent publication.

In the present case, we used both large-area scans yielding monochromatic images $I(\lambda, x,y)$ with spatial coordinates
$(x,y)$, and spatio-temporal time series giving $I(\lambda, x,t)$. Equation
(\ref{eq1}) is insensitive to this difference. We note that $I(\lambda,
x,y)$ and $I(\lambda, x,t)$ are equivalent only in a statistical sense if they
both cover a sufficient large number of periods, or more general, variation
times. In a time series, all phases of the variations are seen at a fixed
location, whereas for a large area map the phases are sampled at different locations in a random state. This requires that the spatial maps should be as large as possible to avoid contamination of the correlation matrices by a coherent evolution in small subfields ($\gg$ 5$^{\prime\prime}$), and that the time-series should at least cover some variation cycles ($\gg$ 5 min).
.
\section{Correlation matrices from observations\label{obsmat}}
To obtain correlation matrices for the various spectral lines, we chose
several observations of quiet Sun (QS) areas at disc centre taken { between 2006 and 2009}. The QS data were taken with slit-spectrograph systems, where the 1-D
slit is stepped across the solar image to obtain a 2-D field of view (FOV). Different spatial locations in the
scanning direction are thus sampled at different times. The observation data
were acquired with the spectropolarimeters POLIS \citep[][396 nm,
630nm]{beck+etal2005b}, TIP \citep[][1083 nm, 1565 nm]{collados+etal2007}, and
a spectroscopic setup using the main spectrograph of the German Vacuum Tower Telescope (VTT) for \ion{Ca}{II} IR 866 nm. Simultaneously with the \ion{Ca}{II} IR 866 nm line, also \ion{Ca}{II} H intensity spectra were recorded for the investigation of cross-correlations; these spectra were, however, not used here. 
{ For \ion{Ca}{II} IR 854 mn, we have two data sets. The first was recorded at the VTT with a similar setup as for \ion{Ca}{II} IR 866 nm. The second} data set of \ion{Ca}{II} IR 854 nm was taken with the IBIS spectrometer \citep{cavallini2006} of the Dunn Solar Telescope (Sac Peak/NSO) in spectropolarimetric mode. IBIS is a Fabry-Perot-Interferometer based 2-D spectrometer. We selected
observations that covered large FOVs on or near disc centre, { and one time-series}; the \ion{Ca}{II} IR 854 nm observation { with IBIS} is the only exemption ({ no QS, but} a pore off disc centre). Appendix \ref{appa} shows the observed FOVs; more details on each observation are given in Table \ref{tabobs}. 

\begin{figure*}
\resizebox{9cm}{!}{\includegraphics{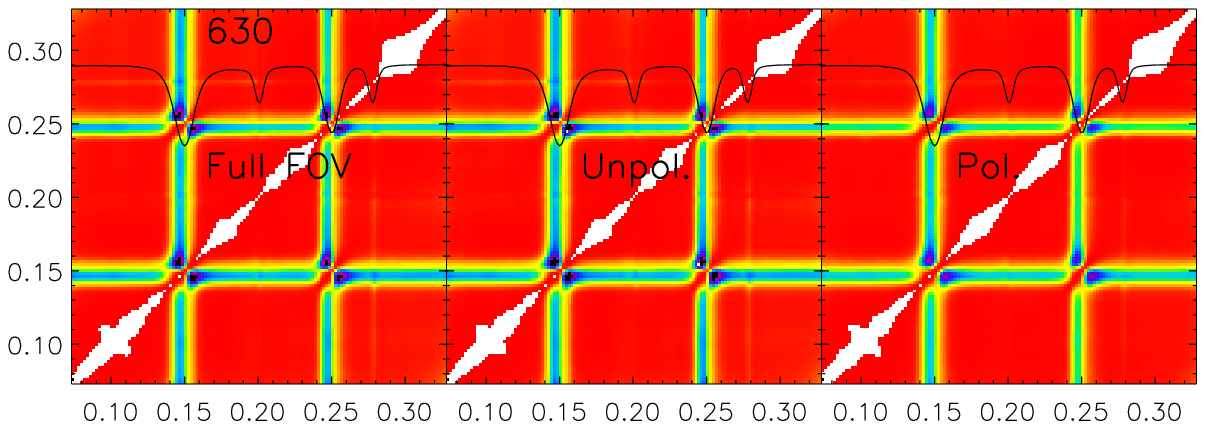}}\hspace*{.75cm}\resizebox{7cm}{!}{\includegraphics{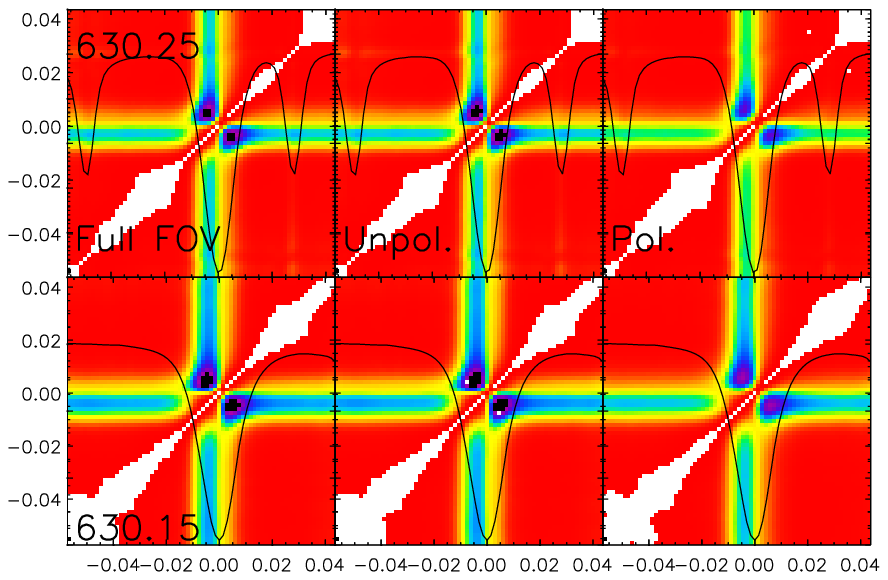}}\hspace*{.2cm}\includegraphics{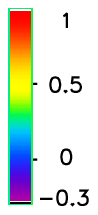}\\$ $\\
\resizebox{9cm}{!}{\includegraphics{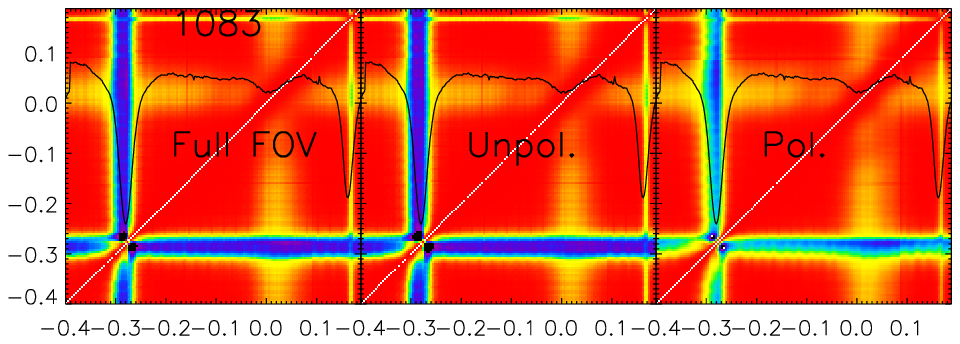}}\hspace*{.75cm}\resizebox{6cm}{!}{\includegraphics{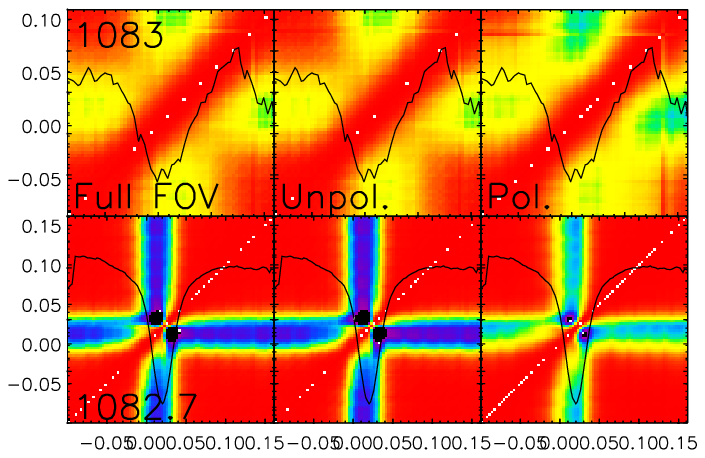}}\hspace*{.2cm}\includegraphics{zbar.ps}\\$ $\\
\resizebox{9cm}{!}{\includegraphics{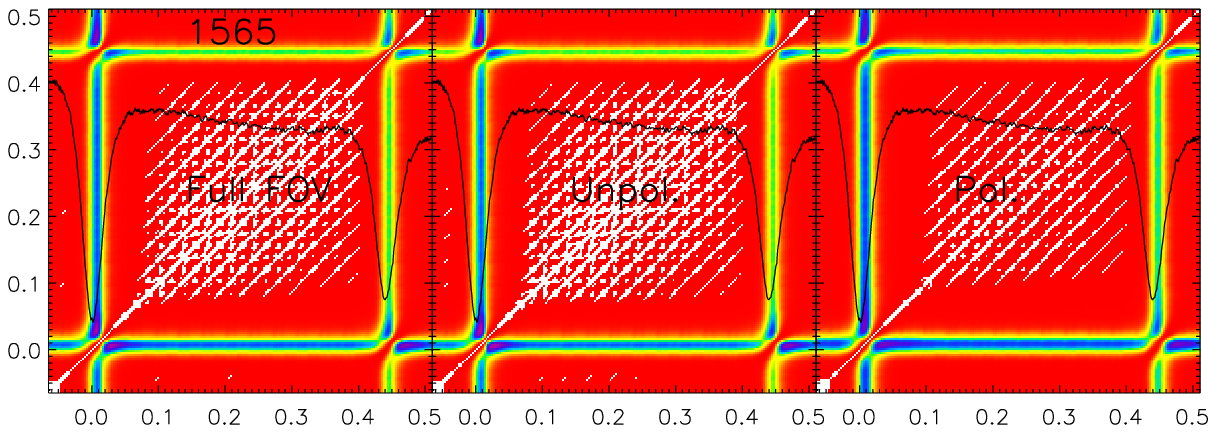}}\hspace*{.75cm}\resizebox{6cm}{!}{\includegraphics{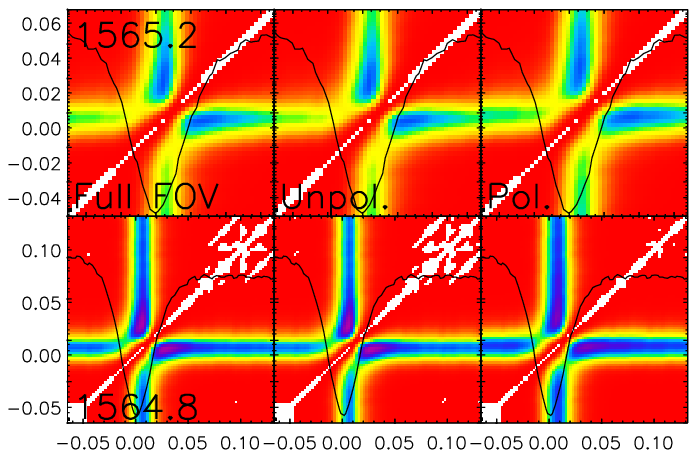}}\hspace*{.2cm}\includegraphics{zbar.ps}\\
\caption{Same as Fig.~\ref{fig3} for 630\thinspace nm, 1083\thinspace nm and 1565\thinspace nm. {\em Top} and {\em bottom row} of the magnification show 630.25\thinspace and 630.15\thinspace (630), \ion{He}{I} 1083\thinspace nm and \ion{Si}{I} 1082.7\thinspace nm (1083), and 1565.2\thinspace and 1564.8\thinspace (1565), respectively. The wavelengths for the left graphs are $\lambda$-630 nm, $\lambda$-1083\thinspace nm and $\lambda$-1564.85\thinspace nm, respectively.\label{fig6}}
\end{figure*}

POLIS, TIP, and IBIS measured the Stokes vector in magnetic sensitive
spectral lines, which additionally gives the option to localize magnetic fields, and to avoid or include the respective positions in the analysis. To investigate the influence of photospheric magnetic fields on the correlation matrices, we created masks of { magnetic}
field locations for each data set. We integrated the absolute circular
polarization signal, $|V|$, in wavelength, and then set a variable threshold
depending on the integration time and magnetic sensitivity of the observed
spectral lines. Locations with polarization signal below (above) the threshold
were assumed to be nearly field-free (magnetic) and are labeled
``unpolarized/unpol.'' (``polarized/pol.'') in the following. As third sample,
we also calculated the correlation matrix using the { full} FOV, which is better
suited for comparisons when no mask of field locations can be made. Appendix
\ref{appa} shows also the Stokes $V$ maps and the masks of each
observation. For the observations of \ion{Ca}{II} IR at 866 nm, no
polarimetric measurements were available. We did not try to define a mask of
locations with presumably magnetic fields from the line core intensity map
(see Fig.~\ref{cairfov}), but only calculated the correlation matrix of the
full FOV in that case. { For the observations of \ion{Ca}{II} IR at 854 nm  taken at the VTT in 2009, we have not yet aligned the data to the simultaneous polarimetric observations with TIP and POLIS, thus also no mask is available.}

\begin{figure}
\resizebox{8.8cm}{!}{\hspace*{.5cm}\includegraphics{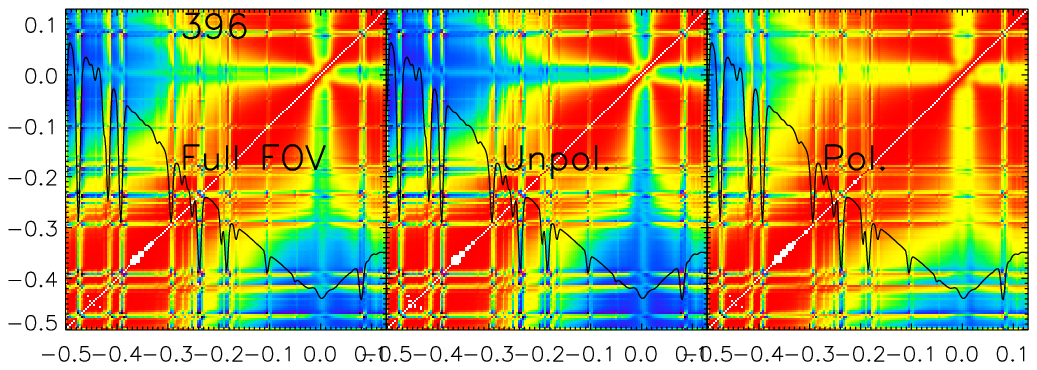}\hspace*{.2cm}\includegraphics{zbar.ps}}\\$ $\\
\resizebox{8.8cm}{!}{\hspace*{-.2cm}\includegraphics{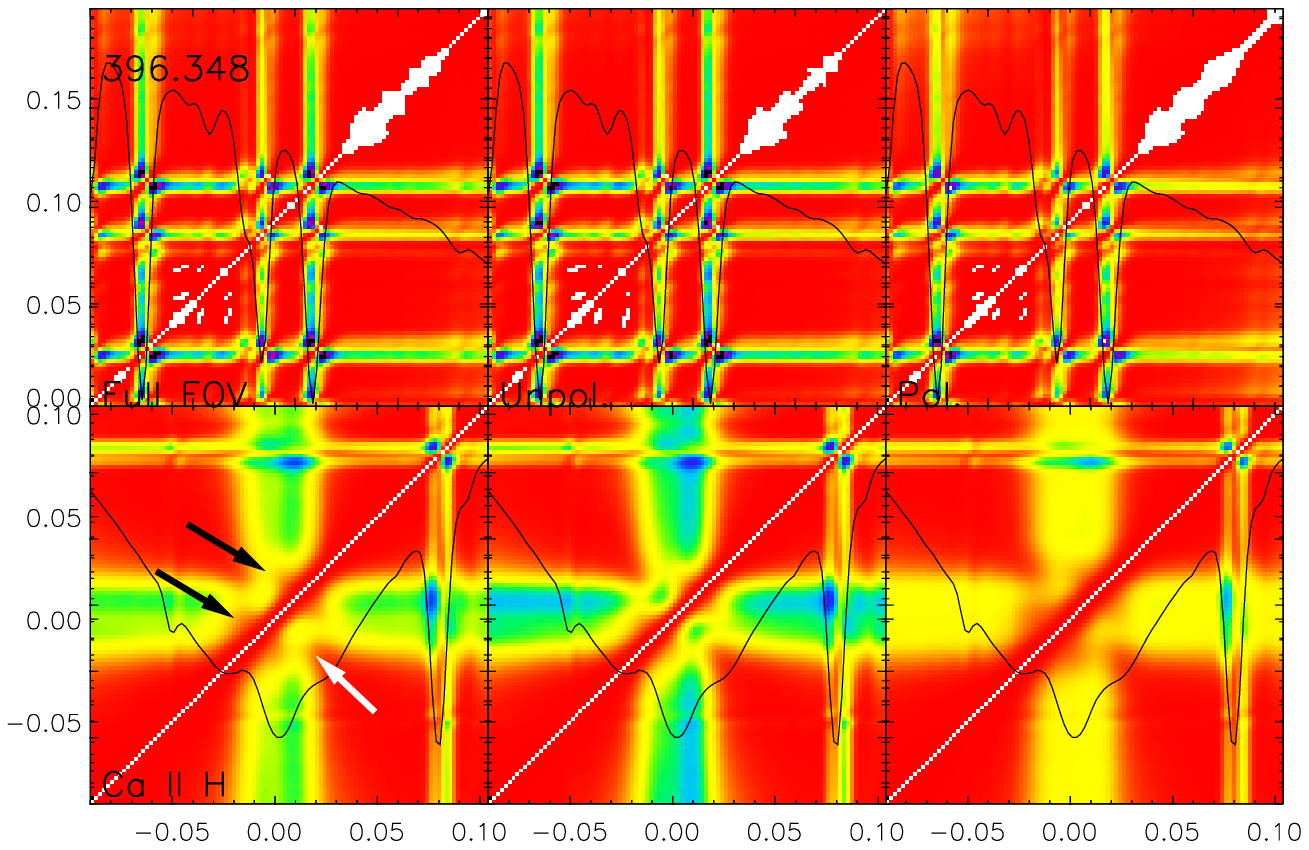}\hspace*{.2cm}\includegraphics{zbar.ps}}\\
\caption{Correlation matrices of the 396\thinspace nm range. {\em Top}: full
  observed wavelength range. {\em Bottom graph}: magnification of line cores
  of some blends { (\ion{Fe}{I} at 396.455\thinspace nm, \ion{Ti}{I} at
    396.427\thinspace nm, and \ion{Cr}{I} at 396.369\thinspace nm)} ({\em
    top}) and \ion{Ca}{II} H 396.85\thinspace nm ({\em bottom}). The matrices
  of the three samples made from the FOV (full FOV, unpol(arized) and
  pol(arized)) are aligned {\em left to right} in each plot. A line profile is
  overplotted as {\em black line}. Wavelengths are $\lambda$-396.85\thinspace
  nm. { {\em White color} corresponds to a correlation value of 1.}\label{fig3}}
\end{figure}

\begin{figure}
\centerline{\resizebox{3.cm}{!}{\includegraphics{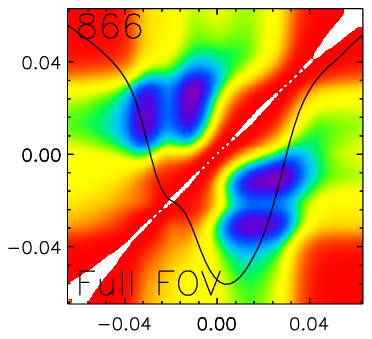}}\hspace*{.2cm}\includegraphics{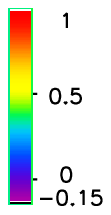}}$ $\\
\centerline{\resizebox{3.cm}{!}{\includegraphics{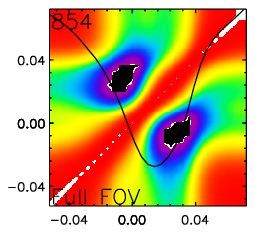}}\hspace*{.2cm}\includegraphics{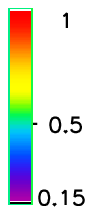}}$ $\\$ $\\
\centerline{\hspace*{.5cm}\resizebox{7.cm}{!}{\includegraphics{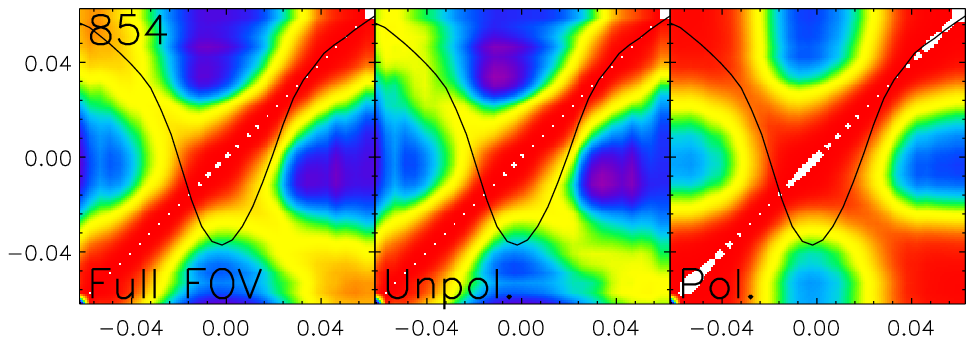}}\hspace*{.2cm}\includegraphics{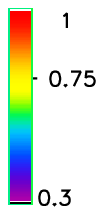}}$ $\\
\caption{{\em Top}: Correlation matrix for \ion{Ca}{II} IR 866 nm for the full
  FOV. Wavelengths are $\lambda$-866.215\thinspace nm. { {\em Middle}: correlation matrix for \ion{Ca}{II} IR 854.2 nm for the full FOV (slit spectrograph data, disc centre).} {\em Bottom}: correlation matrices for \ion{Ca}{II} IR 854.2 nm { (IBIS data, off centre in an active region)}. Wavelengths are $\lambda$-854.215\thinspace nm.\label{caircorrmat}}
\end{figure}

Figures \ref{fig6}, \ref{fig3} and \ref{caircorrmat} show the correlation
matrices, either obtained by averaging the matrices of several observations in
each wavelength range, whenever multiple suited data sets were available, or
derived from individual observations (\ion{Ca}{II} IR lines). Each figure
shows the correlation matrix for the full spectral range covered in each
observation, and additionally a magnified view of individual spectral lines
for the TIP and POLIS data sets. Besides for \ion{Ca}{II} IR 866 nm { and the QS data set of \ion{Ca}{II} IR 854 nm}, each time
the three samples full FOV, unpolarized and polarized are shown {\em left to
  right} in each plot; for \ion{Ca}{II} IR 866 nm { and the QS data set of \ion{Ca}{II} IR 854 nm}, only the matrix of the full
FOV is shown ({ {\em top} and {\em middle panel} of} Fig.~\ref{caircorrmat}). We did not find significant changes of the correlation
matrices with different integration times for photospheric lines, even if for
example for 1565 nm integration times between 5 and 30 secs were used in the
various observations (cf.~Table \ref{tabobs}). { The spectra at 1.56 micron taken with the TIP II camera \citep{collados+etal2007} have a pattern of spectral fringes \citep[see, e.g.,][]{beck+rezaei2009} that produces high correlation on stripes parallel to the diagonal that are not of solar origin.} We first calculated correlation matrices for a time-series with fixed slit separately, but they were qualitatively identical to those from large-area maps and thus were simply included in the
averaging. The \ion{He}{I} 1083 nm line shows nearly no absorption in QS
conditions and thus yields little structure, but has been displayed for
completeness.

The correlation matrices for { the intensity at various} wavelengths and spectral lines forming in
photospheric atmosphere layers (continuum intensity, 630.15 and 630.25 nm, \ion{Si}{I} 1082.7 nm, 1564.8 and 1565.2 nm, the line blends in the wing of \ion{Ca}{II} H) are very similar regardless of wavelength in near-UV, visible or
near-IR. Correlations between { the intensities at} continuum wavelengths are always close to 1. Photospheric spectral lines show { a reduced correlation up to} anti-correlation, if { the intensities at} wavelengths in or close to the line core are correlated with { the intensities at} continuum wavelengths.  { Wavelengths in the red and blue wing are not equivalent: the reduction of correlation between core and red wing is always stronger than for the blue wing}. The absolute value of { the} correlation is smaller for 1565.2 nm and 630.25 nm than for 1564.8 nm or 630.15 nm. We ascribe this fact to the line
depth of the respective spectral line. 

{ The correlation of the line-core intensity of a photospheric line with the intensity at close-by wavelengths ($r_{\lambda_1 \lambda_1\pm\Delta\lambda}$) gives roughly the same pattern as the correlation with the intensity near another line core ($r_{\lambda_1 \lambda_2\pm\Delta\lambda}$) each time $\lambda_2$ corresponds to a photospheric line-core wavelength, doubling the pattern of the stripes of reduced correlation for each spectral line present inside the wavelength range (Fig.~\ref{fig6}). The correlation changes only slightly from line to line, which is due to the fact that the respective photospheric line pairs in the visible and near-IR wavelength ranges (1564.8/1565.2 nm, 630.15/630.25 nm) have similar formation heights \citep{cabrera+bellot+iniesta2005}. If the formation height differs strongly, like for \ion{Si}{I} 1082.7 n and \ion{He}{I} 1083 nm or the line core and all the line blends of \ion{Ca}{II} H, $r_{\lambda_1 \lambda_1\pm\Delta\lambda}$ and $r_{\lambda_1 \lambda_2\pm\Delta\lambda}$ show little similarity. }

With respect to the different samples made in each FOV (full FOV, unpol., pol.) only small changes are seen. A clear trend is that for the polarized sample the width of the { stripes with reduced correlation} related to the line cores is slightly larger, clearly seen only for 630.25 nm and 1564.8 nm. This is due to the splitting of the lines inside magnetic fields, leading to multiple spectral components in the intensity profile\footnote{Tested on synthetic spectra.}. The correlation values between line core(s) { intensities} and { the intensities at} continuum wavelengths are always higher in the polarized sample than in the unpolarized sample; this is best seen for \ion{Ca}{II} H and \ion{Si}{I} 1082.7 nm where the graphs of ``unpol'' and ``pol'' are markedly different in the { value} of the correlation coefficient. The higher correlation could be due to the field-strength dependent shift of the optical depth scale. The intensity will be enhanced at {\em all} wavelengths on locations with magnetic fields, which itself will not increase the correlation due to the subtraction of the average values in Eq.~(\ref{eq1}). But if additionally a dependence of the intensity enhancement on the field strength exists, $I(\lambda_i,x,y)$ and $I(\lambda_j,x,y)$ will scale in the same way. For example, for all locations $(x,y)$ with stronger (weaker) than average magnetic fields, $dI_i = I(\lambda_i,x,y)-\langle I(\lambda_i)\rangle$ and $dI_j=I(\lambda_j,x,y)-\langle I(\lambda_j)\rangle$ will be greater (smaller) than zero at the same time, increasing the fraction of locations with a positive correlation ($dI_i\cdot dI_j >0$). 

The structures in the correlation matrices of all chromospheric \ion{Ca}{II}
lines are significantly different from the photospheric cases. For
\ion{Ca}{II} H (Fig.~\ref{fig3}), the correlation of { the intensities at} wavelengths in or close to the Ca line core with { that at} wavelengths separated by around 0.3 nm gives a { reduced correlation over an extended wavelength range}; the { reduction of correlation} is not restricted to only a small stripe near the line core like for the photospheric lines. Around the diagonal, a band of high correlation exists, whose width changes with wavelength, increasing from the \ion{Ca}{II} H line core until around 0.2 nm wavelength separation, decreasing from 0.2 nm until 0.3 nm, and then increasing again. The effect is most pronounced for the polarized case, where the high correlation extends over almost the full wavelength range for wavelengths outside of spectral lines. 
\begin{figure*}
\centerline{\hspace*{1cm}\resizebox{4.cm}{!}{\includegraphics{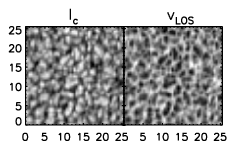}}\hspace*{1.cm}\resizebox{3.cm}{!}{\includegraphics{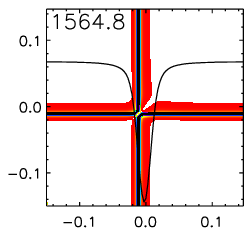}}\hspace*{1cm}\resizebox{3.cm}{!}{\includegraphics{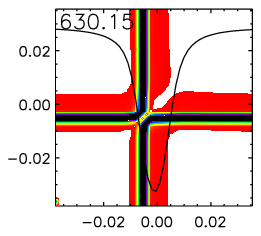}}\hspace*{1cm}\resizebox{3.cm}{!}{\includegraphics{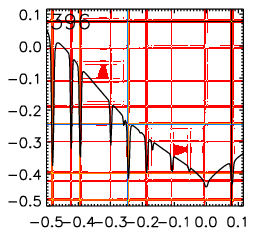}}\hspace*{1cm}\includegraphics{zbar.ps}}$ $\\$ $\\
\centerline{\hspace*{1cm}\resizebox{4.cm}{!}{\includegraphics{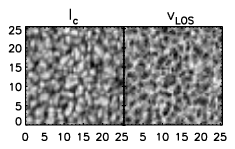}}\hspace*{1cm}\resizebox{3.cm}{!}{\includegraphics{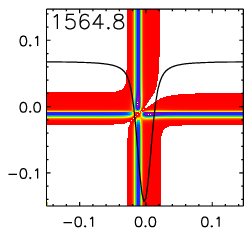}}\hspace*{1cm}\resizebox{3.cm}{!}{\includegraphics{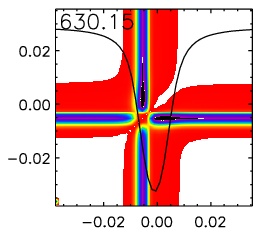}}\hspace*{1cm}\resizebox{3.cm}{!}{\includegraphics{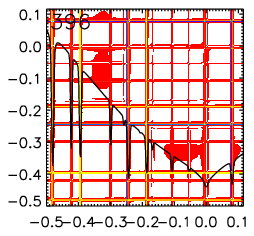}}\hspace*{1cm}\includegraphics{zbar.ps}}$ $\\
\caption{{ Simulation of granulation and resulting correlation matrices. {\em 1st column}: continuum intensity $I_c$ and LOS line-core velocity maps; tick marks are in arcsec. {\em 2nd to 4th column}: correlation matrices for 1564.8 nm, 630.15 nm, and the full Calcium line profile. {\em Top row}: velocity directly proportional to $I_c$. {\em Bottom row}: same as before, but with a random velocity added.}\label{gran_sim}}
\end{figure*}

The photospheric line blends in the Ca
line wing show the same pattern with the red/blue asymmetry near their line
core as the other photospheric lines, but the \ion{Ca}{II} H line core shows a
stronger structured pattern. It consists of two stripes of { reduced} correlation starting from the line core like for the photospheric lines, but at around
0.15 nm from the line core a small area of { reduced} correlation in the shape of a half-circle is visible that is missing for the other lines (white arrow in
Fig.~\ref{fig3}). The feature connects the two stripes of { reduced} correlation going to red and blue wing, respectively. Centred on around $\pm 0.02$ nm
from the line core, two small squares of high correlation can be seen (black
arrows in Fig.~\ref{fig3}), directly above (left) of the lower
half-circle. The blue/red asymmetry of the photospheric lines is missing
completely, both wings seem to be equivalent for the \ion{Ca}{II} H line
core. { \ion{Ca}{II} IR at 866 nm ({\em top panel} of Fig.\ref{caircorrmat}) also shows a more complex pattern compared to photospheric lines. It exhibits two patches of reduced correlation} at around +0.03 nm from the core{, presumably caused by a
  strong \ion{Fe}{I} line blend at this wavelength{, but blue and red wing are equivalent without asymmetry\footnote{{ Only the wavelength range near the line core is shown in Fig.~\ref{caircorrmat}.}}}. The correlation matrix for the QS data of \ion{Ca}{II} IR at 854 nm ({\em middle panel} of Fig.~\ref{caircorrmat}) is fairly similar to that of the 866 nm line, but shows only a single patch of reduced correlation. The structure seen in the off centre observations of \ion{Ca}{II} IR at 854 nm is quite different.} The { stripes of reduced correlation} to the red and blue of the line core { do not} connect  directly { like in the QS data}; { especially} for the polarized sample a high correlation is found between the branches from core to
blue and red. { We suggest that the strong variation of the correlation matrix for one and the same spectral line is due to both the solar structure of the observed FOV (cf.~Fig.~\ref{cairfov}) and the off centre position, similar to the differences in the temperature correlations later on in Sect.~\ref{secttemp}}. In total, the photospheric lines present a simple structure in the correlation matrices dominated by the red/blue asymmetry, whereas all chromospheric lines show a more detailed fine-structure.
\section{Numerical and (semi-)analytical spectra\label{theomat}}
For comparison to the observed correlation matrices, we used spectra obtained from { four} approaches differing in the sophistication of the method used for the generation of the spectra.
\subsection{{ Granulation simulation}\label{gransimm}}
{ The convective energy transport and the resulting granulation pattern dominates the spatial distribution of the continuum intensity in the solar photosphere, and also all layers up to around 300-400 km above log $\tau_{500 {\rm\, nm}}$ = 0 that are relevant for the formation of photospheric spectral lines. To estimate the contribution of granulation to the wavelength correlation matrices, we created a synthetic data set that includes only granulation, and no waves of any kind. We used a subsection of the continuum intensity map of a long-exposed quiet Sun observation on disc centre obtained with the TIP instrument at 1.56 micron \citep{beck+rezaei2009} to define the intensity pattern in a 25 arcs$^2$ area (see {\em left column} of Fig.~\ref{gran_sim}). We defined the LOS velocity to be inverse proportional to the intensity by 
\begin{equation}
v_{LOS}(x,y) = -\,\Delta I_c(x,y) \cdot 20000\, {\rm ms}^{-1} \, ,\label{eq_velo}
\end{equation}
which yields velocities in around a $\pm$ 1 kms$^{-1}$ range, since $\Delta I_c(x,y) = (\, I_c(x,y)-\langle I_c \rangle\,)$ lies between around 0.95 and 1.05 for the near-IR observations. 

The temperature stratification was defined using a modified version of the Harvard Smithsonian Reference Atmosphere model \citep[HSRA,][]{gingerich+etal1971}. We used an optical depth range from log $\tau$=1.4 to log $\tau$=-6. Up to log $\tau$=-4, we used the values of the original HRSA model, but in the layers above we substituted the chromospheric temperature rise in the HSRA with the corresponding values of the Holweger-Mueller model \citep[HOLMUL,][]{holweger+mueller1974}.  The Holweger-Mueller model gives to first order an atmosphere in radiative equilibrium without a chromospheric temperature rise. The main motivation for this choice was that otherwise in a LTE synthesis Ca spectra with (unobserved) large single-peaked emission in the line core result, but \citet{rezaei+etal2008} also showed that the solar chromosphere partially reverts to such a low-energy state in the absence of shock waves. The temperature stratifications of the two models were smoothly connected by adding the difference $\Delta T_{{\rm log}\tau =-4} = T({\rm HSRA})_{{\rm log}\tau =-4}-T({\rm HOLMUL})_{{\rm log}\tau =-4}$ to the values of the HOLMUL model. The gas and electron density above log $\tau$=-4 were derived by an extrapolation of the exponential decrease of the HSRA model below log $\tau$=-4 to the rest of the optical depth range. The atmosphere was not put to hydrostatic equilibrium, as only a qualitative guess was desired.

For the granulation simulation, the temperature on each point $(x,y)$ was additionally modified by multiplying the temperature stratification with 1+$\Delta I_c(x,y)$, yielding something like $\pm 300$ K variation at continuum forming layers. Given temperature stratification and LOS velocity, we synthesized spectra for the 1.56 micron range, 630 nm range, and \ion{Ca}{II} H, respectively, in LTE with the SIR code \citep{cobo+toroiniesta1992}. The resulting correlation matrices for 1564.8 nm, 630.15 nm, and the \ion{Ca}{II} H line are shown in the {\em top row} of Fig.~\ref{gran_sim} from {\em left to right}. They show stripes of, in this case, not only reduced but anti-correlation starting at the line cores like in the observations, but do not exhibit the red/blue asymmetry. The correlation values are much lower than for the observations and reach down to around -0.9 near the very line core of any photospheric line. For \ion{Ca}{II} H, the correlation basically is unity over the complete wavelength range outside the photospheric line blends. The asymmetry for photospheric lines thus does not originate from the presence of hot upflows and cold downflows, if a perfect linear relation between intensity and velocity is assumed. 
\begin{figure}
\centerline{\resizebox{6.cm}{!}{\includegraphics{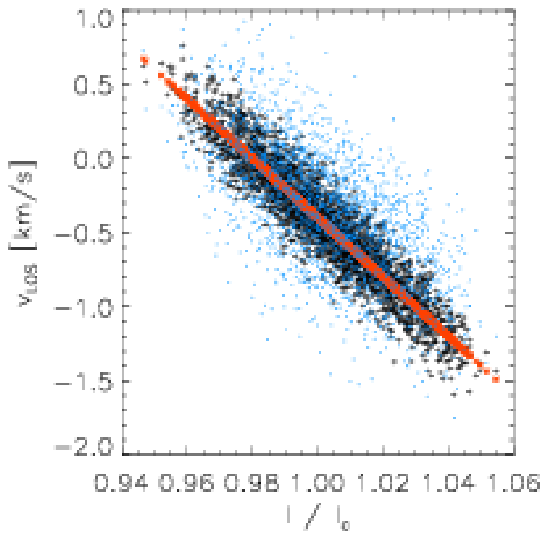}}}$ $\\
\centerline{\hspace*{1.cm}\resizebox{4.cm}{!}{\includegraphics{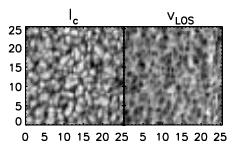}}}$ $\\
\caption{{ {\em Top}: scatterplot of LOS line-core velocities vs continuum intensity $I_c$. {\em Red}: velocity derived from Eq.~(\ref{eq_velo}). {\em Black}: velocity from Eq.~(\ref{eq_velo}) with random variation added. {\em Blue}: observed velocity. {\em Bottom}: comparison of observed intensity and LOS velocity.}\label{fig_velo}}
\end{figure}

As second test, we used the same setup as above for the temperature and velocity, but added a random Gaussian variation of 150 ms$^{-1}$ variance to the velocities derived from Eq.~(\ref{eq_velo}), and calculated again the correlation matrices from the corresponding synthetic spectra ({\em lower row} in Fig.~\ref{gran_sim}). The correlation matrices then show the a red/blue asymmetry for the photospheric lines, albeit the correlation values are still much lower than for the observations. The matrix of \ion{Ca}{II} H does not change noticeably and has no resemblance to the observed correlation matrix. 
\begin{figure}$ $\\$ $\\
\centerline{\resizebox{8cm}{!}{\includegraphics{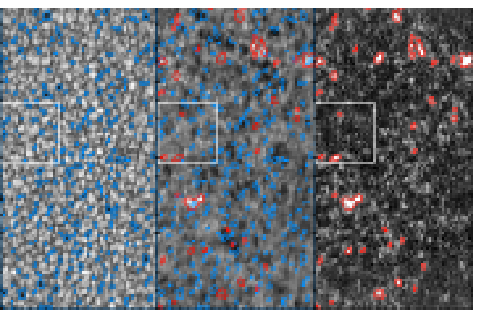}}}$ $\\$ $\\$ $\\
\centerline{\resizebox{8cm}{!}{\includegraphics{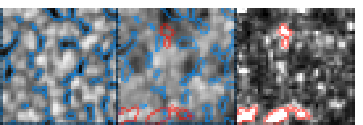}}}$ $\\$ $\\$ $\\
\centerline{\resizebox{8cm}{!}{\includegraphics{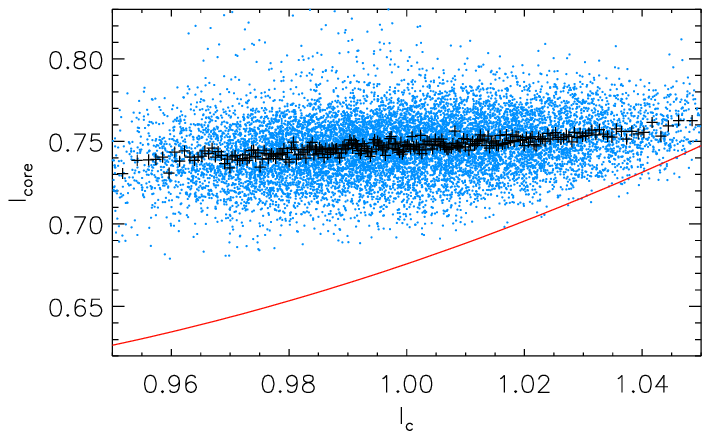}}}
\caption{{ Relation between continuum intensity ($I_c$) and line core intensity ($I_{core}$). {\em Top, left to right}: $I_c$, $I_{core}$, polarization degree. {\em Red contours} trace high polarization degree, {\em blue contours} low $I_c$. {\em Middle}: magnified view of the {\em white rectangle} marked above. Tick marks are in arcsec. {\em Bottom}: scatterplot of $I_c$ and $I_{core}$. {\em Red line}: granulation simulation, {\em blue dots}: observation, {\em black crosses}: same after binning (see text for details).\label{icicorefig}}}
\end{figure}
\begin{figure}
\centerline{\resizebox{6.cm}{!}{\includegraphics{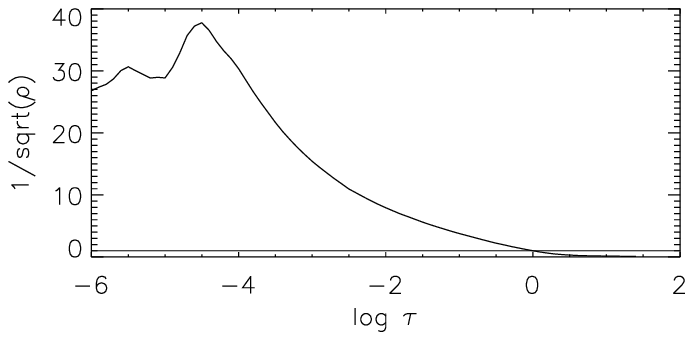}}}
\centerline{\resizebox{5.8cm}{!}{\includegraphics{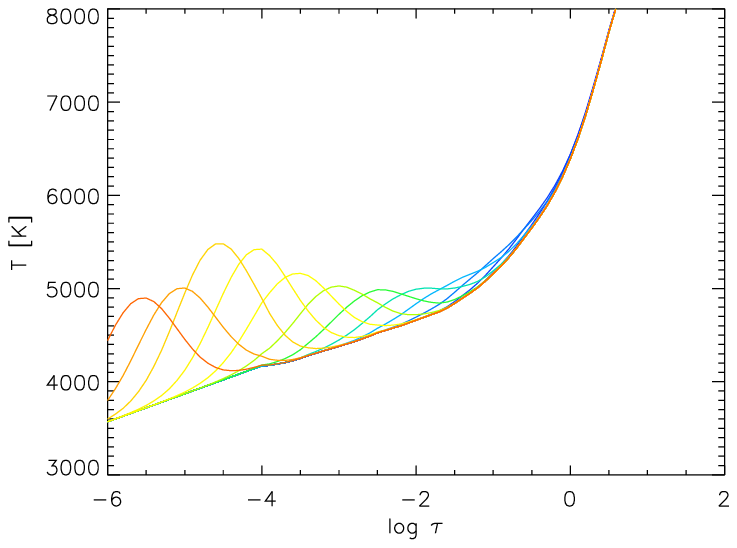}}}
\centerline{\resizebox{5.8cm}{!}{\includegraphics{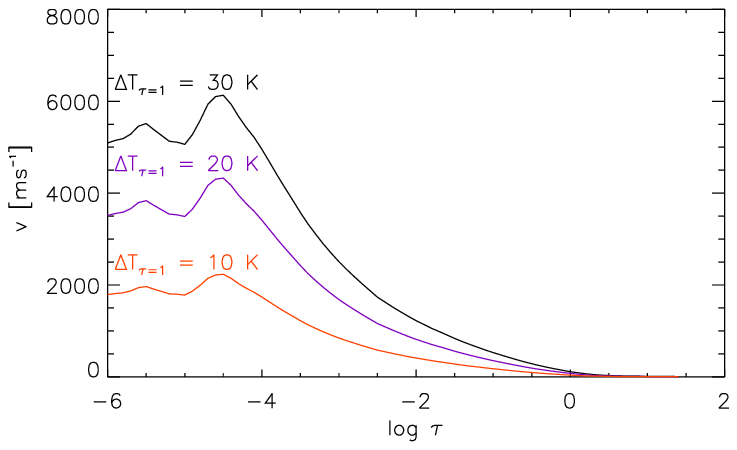}}}
\caption{{\em Top left}: scaling law for the amplitudes of perturbations with optical depth. {\em Top right}: some of the temperature stratifications for the 20 K run. {\em Bottom}: velocity perturbation for three different temperature perturbations.\label{fig8}}
\end{figure}

Figure \ref{fig_velo} compares the velocities of the two synthetic approaches with the actually observed line-core velocity of the 1564.8 nm line. The scatter of the observed velocity ({\em blue}) around the linear relationship ({\em red}) defined by Eq.~(\ref{eq_velo}) is about three times as large as the tested 150 ms$^{-1}$ variance ({\em black}). This suggest that the asymmetry between red and blue wing for wavelengths near the line core of photospheric lines is not due to the granulation pattern of hot upflows and cold downflows, but the deviation from the purely convective source of the velocity field, i.e., the (magneto-)acoustic waves that usually have no strong signature in intensity at photospheric levels. The observed correlation matrix of the chromospheric \ion{Ca}{II} H seems to be fully unrelated to a pure granulation pattern. 

Another possible source of the asymmetry in the correlation matrix of photospheric lines could be the temperature (or intensity) contrast inversion between granules and intergranular lanes above the photosphere \citep[``{\it reverse granulation}'', see, e.g.,][]{rutten+etal2004a}. To find out whether this effect influences the correlation matrix, we investigated the relation between continuum intensity $I_c$ and the line-core intensity $I_{core}$ of the 1564.8 nm line. Figure \ref{icicorefig} shows the corresponding maps for the full FOV of the quiet Sun observations, together with a map of the polarization degree of the spectra. The line-core intensity was defined as the minimum intensity value in the spectral region of the 1564.8 nm line. The line-core intensity of the low-forming 1564.8 nm line \citep{cabrera+bellot+iniesta2005} still reflects the granulation pattern with a small positive correlation coefficient of 0.26 with respect to  $I_c$, i.e., $I_{core}$ is not yet influenced by the reverse granulation pattern. Mainly where the polarization degree is high, $I_{core}$ and $I_c$ exhibit inverse patterns. The relation between $I_{core}$ and $I_c$ shows a steeper slope for the granulation simulation than for the observations ({\em bottom panel} of Fig.~\ref{icicorefig}), but this could be due to the stray light contribution to the observed spectra. The binned values were derived like in \citet{beck+etal2007} by averaging the pairs $(I_c, I_{core})$ over bins in $I_c$, where the plotted points correspond to $(\langle I_c^i \rangle, \langle I_{core}^i \rangle)$. They correspond to the center of gravity of the $ I_{core}$ distribution as function of $I_c$. From the positive correlation of $I_c$ and $I_{core}$ and the correlation matrix of the granulation simulation, we think that we can exclude the granulation pattern as source of the asymmetry for photospheric line, even if, e.g., the lines at 630 nm form in slightly higher layers closer to the reverse granulation pattern than the near-IR lines.}
\subsection{Simplified 1-D LTE synthesis { of wave propagation}}
{ In the next attempt to reproduce the observed correlation matrices, synthetic spectra were generated by assuming an upwards propagating wave that creates a temperature perturbation of the modified HSRA model as defined above.} The perturbation was modeled as a Gaussian with a width of
$\sigma$=0.4 units in log $\tau$, and an amplitude $A$ that was scaled up with
decreasing optical depth. We scaled the amplitude with the electron pressure
that is related to gas density using $A$=$A_0$/$\sqrt{p_{el}}$ (see {\em top
  panel} of Fig.~\ref{fig8}). We normalized the scaling law for the
amplitude to 1 at log $\tau$=0 to be able to prescribe exact initial values at
this optical depth. We then synthesized spectra with the SIR code
 for the Gaussian perturbation moving through all
75 points of the optical depth grid (log $\tau$ +1.4 to -6;  sampling 0.1
units of log $\tau$), only modifying the temperature. The {\em middle
  panel} of Fig.~\ref{fig8} shows some of the temperature stratifications that
result for an initial perturbation of 20 K at log $\tau$=0. The
propagation speed of the perturbation is constant in the optical depth scale,
not in geometrical height, and without adding a corresponding velocity
perturbation no Doppler shifts are induced.
\begin{figure}
\resizebox{8.8cm}{!}{\hspace*{.5cm}\includegraphics{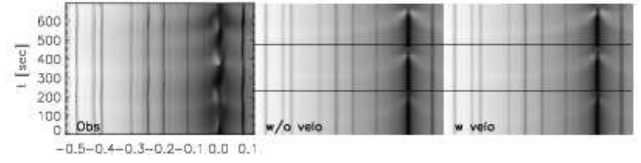}}$ $\\
\caption{Comparison of observed ({\em left}) and synthetic 1-D LTE spectra
  without velocities ({\em middle}), and with velocities included ({\em
    right}). { The three runs with 10,20,30 K initial perturbation are
    shown contiguously from {\em bottom to top} in the simulated spectra.}\label{fig23}}
\end{figure}

Figure \ref{fig23} compares a sample of observed profiles at a fixed spatial location over a time of around 600 secs ({\em left}) with the synthetic spectra resulting when propagating an initial temperature perturbation of 10,20,30 K at log $\tau$ = 0 through the optical depth grid ({\em middle}{ , from {\em bottom to top} for 10,20,30 K}). We synthesized the \ion{Ca}{II} H line with all of its line blends for which atomic parameters were at hand. The photospheric lines at 630 nm were also synthesized to be conform with the standard wavelengths of POLIS.

The observed profiles have been stretched along the y-axis to match the
evolution ``speed'' of the synthetic ones\footnote{{ The cadence of the observations was around 20 seconds, yielding fewer spectra than in the wave simulation.}}. The brightenings originating in the line wing and culminating in strong emission in the core can be clearly traced in the both synthetic and observed spectra. The observed bright grains
increase in intensity from the 1st to the 2nd one like for the synthetic
spectra, but this actually was only by chance; the observations were selected
before the synthetic spectra were calculated. The complex
behavior of the Ca line core in the observations is of course completely
missing in the synthetic spectra due to the lack of velocities and the LTE
assumption. The synthetic spectra are fully symmetric with respect to the Ca
rest wavelength. Using the synthetic spectra for the three emulated ``waves'',
we calculated the wavelength correlation matrices in the same way as for the
observations. We remark that in this case only 3$\times$75 spectra were used,
opposite to the observations with several ten thousands of spectra, but the
synthetic spectra cover all phases of the wave. This correlation matrix,
however, turned out to be fully symmetric also for the photospheric lines,
i.e., the pronounced red/blue asymmetry (cf.~Fig.~\ref{fig6}) was missing.

To improve the agreement with the observations, we then also introduced a
line-of-sight velocity perturbation in phase with the temperature
perturbation. The velocity amplitude was scaled with the same relation as the
temperature perturbation (Fig.~\ref{fig8}, {\em lower panel}). The velocity
was chosen to be positive, i.e., a downflow, corresponding to a temperature
increase by compression of gas during the propagation of an acoustic wave. { This choice of direction for} the velocity was necessary to produce the
correct asymmetry for the photospheric lines, whereas for the \ion{Ca}{II} H
line core it enhances the red emission peak, contrary to what happens in
observations ({\em right panel} of Fig.~\ref{fig23}). { If the velocity was applied with a time-lag, either leading or lagging the temperature increase, the resulting correlation matrix turned out to be fully symmetric again in red and blue line wing.} 
\begin{figure*}
\hspace*{.3cm}\resizebox{5.cm}{!}{\includegraphics{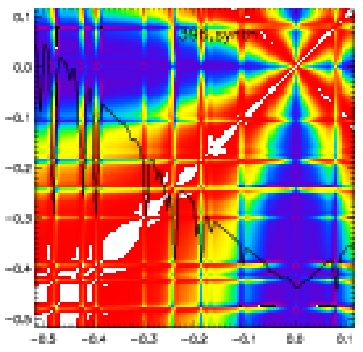}}\hspace*{1cm}\resizebox{3.cm}{!}{\includegraphics{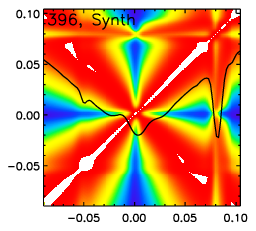}}\hspace*{.3cm}\includegraphics{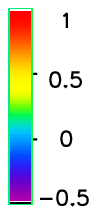}\hspace*{1.cm}\resizebox{3.cm}{!}{\includegraphics{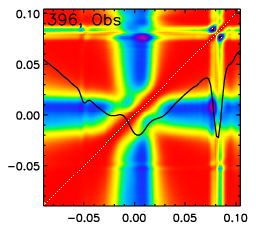}}\hspace*{.3cm}\includegraphics{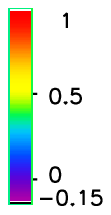}\\$ $\\$ $\\
\caption{Wavelength correlation matrices for \ion{Ca}{II} H from the synthetic
  1-D LTE  spectra including a velocity perturbation. {\em Left}: full
  wavelength range. {\em Middle}: magnified view of the line core. {\em
    Right}: same section  from observations; display range was reduced to
  -0.15 to 1 here{; the lower level of -0.5 applies to the left and middle plot}.\label{fig24}}
\end{figure*}

\begin{figure}
\centerline{\resizebox{4.5cm}{!}{\hspace*{.6cm}\includegraphics{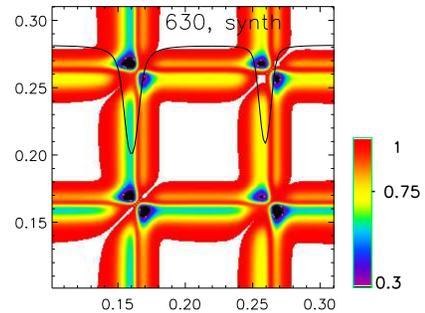}}\hspace*{.3cm}\includegraphics{zbar4.ps}}$ $\\
\caption{Wavelength correlation matrix for 630 nm from the synthetic 1-D LTE spectra including a velocity perturbation.\label{fe_synth}}
\end{figure}

Figures \ref{fig24} and \ref{fe_synth} show the correlation matrices { for \ion{Ca}{II} H and 630 nm, respectively,} obtained from the synthetic spectra including the positive velocity perturbation. They have to be compared with Figs.~\ref{fig6} and \ref{fig3}. Starting with the photospheric 630\thinspace nm lines, the asymmetry between red and blue wing can be reproduced by the assumed positive velocity. { Similar to the observations, the correlation coefficient is lower, when the intensity at the rest wavelength of the 630.15 nm line is correlated with other wavelengths than for the line core of the 630.25 nm line, which is presumably related to the line depth. The difference of the absolute values of the correlation coefficient between observations (minimum below -0.3) and the simulation (minimum +0.3) could be due to the granulation pattern in the observation that contributes anti-correlation for line-core wavelengths (Fig.~\ref{gran_sim}).} For \ion{Ca}{II} H, the global shape of the observed correlation matrix is roughly reproduced, but with several huge differences to the observations. The band of high correlation around the diagonal widens monotonically with wavelength distance to the Ca line core. Near the Ca line core, the half-circle/bridge between the { stripes with reduced correlation} to red and blue wing is missing; directly in the Ca line core, only one square of high correlation exists, located exactly at the rest wavelength of the \ion{Ca}{II} H core.

{ The fine-structure near the very line core of \ion{Ca}{II} H seen in the correlation matrix derived from the observations thus is not reproduced by the simplified wave simulation. A close match, however, is not to be expected in this case, since the processes in the upper atmosphere during the passage of a shock front are very complex. They include large Doppler shifts of the line due to upwards and downwards direct motions, together with strong variations of the intensity near the line core, where the flow directions in front of and behind the shock front can be opposite \citep{carlsson+stein1997}. Macroscopic mass flows in the atmosphere correspond to the correlation with a shifted line profile that yields the same correlation matrix, but displaced in wavelength, if the profile shape is not changed. To test the effect of macroscopic flows in addition to the simulated wave, we took the correlation matrix for the synthetic \ion{Ca}{II} H spectra of the wave simulation, shifted it along the diagonal, and added it to the original un-shifted one. Shifting the correlation matrix by around 5 kms$^{-1}$ and adding it to the original matrix reproduced the features of the observed correlation matrix near the line core, i.e, the double squares of high correlation and connecting bridge between low correlation stripes, but also doubles the pattern due to the photospheric blends in full contradiction to the observations. We thus suggest that any additional macroscopic flows should only be present in the upper atmosphere layers not seen by the photospheric blends, and that the fine-structure in the correlation matrix near the \ion{Ca}{II} H core could be produced by the complex flow pattern near a shock front that passes through the atmosphere.

The two contributions from the granulation simulation and the simplified wave propagation together seem to be able to reproduce the patterns seen in the
correlation matrix of observed photospheric spectra and the global structure of the \ion{Ca}{II} H correlation matrix, albeit not the fine structure near the line core of the chromospheric line.}
\subsection{NLTE 1-D simulations}
As { third} numerical experiment, we used spectra generated from 1-D NLTE simulations done with the WAVE code \citep{rammacher+ulmschneider2003}. The simulations were done similar to \citet{carlsson+stein1997}, with a photospheric piston that excites acoustic waves of various types. The driver of the piston was varied between stochastic excitation and monochromatic waves. Here we use three runs corresponding to stochastic waves with a mechanical energy flux of 5$\cdot 10^7$ erg cm$^{-2}$s$^{-1}$ and 2$\cdot 10^8$ erg cm$^{-2}$s$^{-1}$, and monochromatic waves of 45 sec period with also 2$\cdot 10^8$ erg cm$^{-2}$s$^{-1}$ as energy flux. 
\begin{figure*}
\centerline{\includegraphics{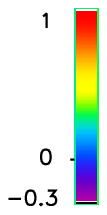}\hspace*{.9cm}\includegraphics{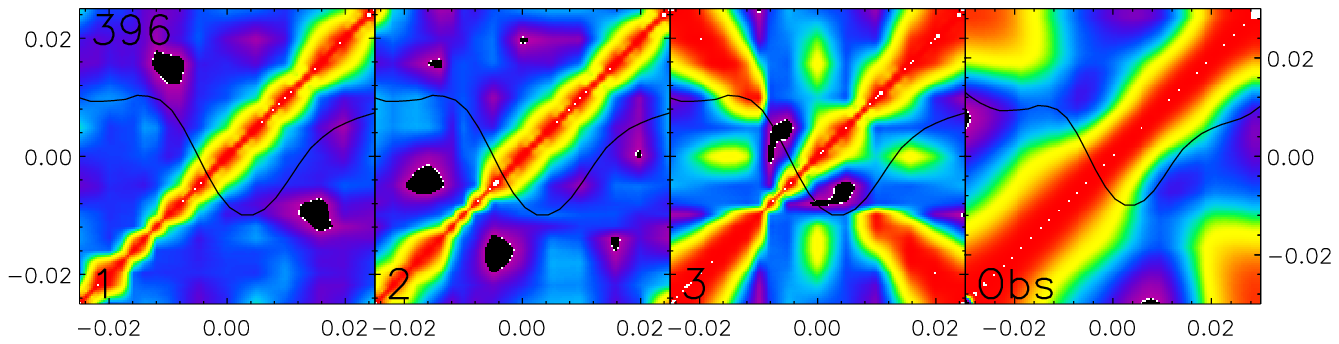}\hspace*{1.cm}\includegraphics{zbar2.ps}}$ $\\
\caption{{\em Left to right}: correlation matrices of the \ion{Ca}{II} H line core from 1-D NLTE calculations for stochastic waves (1,2), monochromatic waves (3), corresponding section of matrix from observations. Wavelength range (-0.03 nm to 0.03 nm) and display range (0.15 to 1) of the observations' matrix are slightly different. Wavelengths are $\lambda$-396.849\thinspace nm.\label{nlte_corr}}
\end{figure*}

The correct calculation of spectra for the chromospheric emission lines requires the use of partial redistribution (PRD). However, a line treatment implementing PRD like in \citet{ulmschneider1994} leads to excessive computation times that cannot be tolerated in time-dependent wave calculations. We thus followed \citet{huenerth+ulmschneider1995} by employing a so-called ``pseudo-PRD'', which computes the line assuming complete frequency redistribution (CRD) but then artificially removes the damping wings from the lines by multiplying the damping parameter in the Voigt function by a factor of 1/100. The spectra were calculated with this method from the full simulation runs only for the very core of \ion{Ca}{II} H in a range of $\pm$0.03 nm around the rest wavelength, but this covers the fine-structure as seen in the observations. Spectral samples of 1500 such profiles were used to calculate the correlation matrices for each piston model.

Figure \ref{nlte_corr} shows the resulting correlation matrices for the runs with stochastic excitation (1 and 2), and the monochromatic case (3). The patterns in the correlation matrices for stochastic excitation do not match at all to the observations ({\em last column} in Fig.~\ref{nlte_corr}). { The matrix for the monochromatic waves, however, fits partly to that of the observed spectra. The central band of high correlation is smaller than for the observations and there exists a cross-like structure of higher correlation values at the rest wavelength, but the high correlation values at $\pm$0.025 nm appear in both observations and simulations. The bridge/half-circle that connects the two bands with reduced correlation starting from the core is present in the simulations as well. Some of the additional fine-structure in the simulations' correlation matrix is missing in the observations, but this actually could also be due to either limitations on the observational side (temporal/spatial/spectral resolution) or the realism of the simulations with a strictly monochromatic permanent driver.}

\begin{figure}
\centerline{\hspace*{.8cm}\resizebox{4.5cm}{!}{\includegraphics{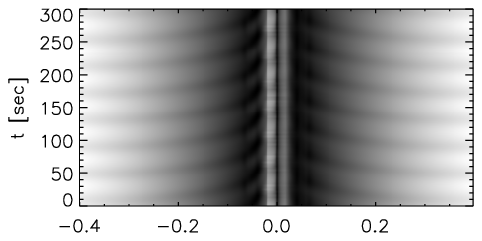}}}$ $\\
\centerline{\hspace*{.8cm}\resizebox{3cm}{!}{\includegraphics{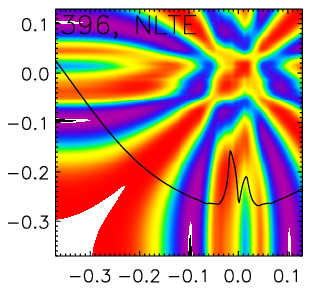}}}$ $\\$ $\\
\centerline{\hspace*{.8cm}\resizebox{3cm}{!}{\includegraphics{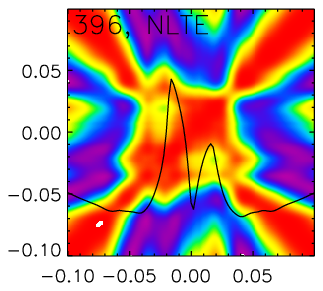}}}$ $\\
\caption{{{\em Top to bottom}}: 1-D NLTE spectra for monochromatic waves, correlation matrix of \ion{Ca}{II} H, magnification of line core. Wavelengths are $\lambda$-396.849\thinspace nm. Display range is $\pm$1.\label{nlte_corr_short}}
\end{figure}

To investigate the correlation matrix for the 1-D NLTE simulations also in the line wings, new spectra were synthesized from the simulation results using the PRD code as described in \citet{ulmschneider1994}. This allows to extend the wavelength range, but unfortunately is quite demanding on computing power. So far, only a set of 100 spectra with 3 seconds { temporal} sampling are available for the case of the monochromatic 45-sec piston (Fig.~\ref{nlte_corr_short}, {{\em top}}). { The spectra were calculated on a non-equidistant wavelength grid with a spectral sampling of around 4 pm in the line wing and 1.5 pm near the line core to reduce the computational effort. We interpolated and re-sampled the spectra to an equidistant sampling of 2 pm for a better match to the observations \citep[1.9 pm sampling at a spectral resolution of 220.000,][]{beck+etal2005b}, and then determined the correlation matrix as before.} The {\em middle} and {{\em bottom}} panel in Fig.~\ref{nlte_corr_short} show the resulting correlation matrix for the full wavelength range and a magnification of the line core, respectively. Similar to the simplified 1-D LTE modeling, the correlation band around the diagonal widens monotonically for increasing distance from the line core, contrary to the observations (cf.~Fig.~\ref{fig3}). The correlation from core to red and blue wing ($r_{0,\lambda_i}$ and $r_{\lambda_i,0}$) now shows positive correlation { for $|\lambda_i| > 0.1$ nm}, contrary to both the observations and the simplified LTE modeling. This { could be} due to the short time span sampled and the strict period of the driver; the correlation of a spectrally un-smoothed version of the spectra { without re-sampling} shows several repeated stripes { of alternating high and low correlation} there. The line core region ({\em right}) matches better to the observations (cp.~to the rightmost panel in Fig.~\ref{fig24}). The two squares of high correlation at $\pm$0.02 nm are present, as well as the bridge between the { stripes with reduced correlation} to red and blue wing, albeit the correlation only drops to zero and does not turn to anti-correlation ($r_{\pm 0.2,\mp 0.2}$).

In total, the 1-D LTE or NLTE calculations are able to reproduce the general
shape of the correlation matrix of \ion{Ca}{II} H in the line wing, whereas
the 1-D NLTE calculation yields a first order match to the very line core. In
both cases, the lower part of the atmosphere is treated similarly as being
permeated by propagating acoustic waves leading to similar spectral patterns
(cp.~the spectra in ~Figs.~\ref{fig23} and \ref{nlte_corr_short}). The line
core region is only reproduced in the NLTE calculations due to the complex
dynamics in the shock fronts that form in the upper atmosphere
\citep[e.g.,][]{rammacher+ulmschneider1992}. The remaining mismatch between
observed and synthetic correlation matrices is presumably due to the
granulation pattern that is absent in the synthetic spectra { of the wave simulations}. 
\begin{figure}
\hspace*{.75cm}\resizebox{2.5cm}{!}{\includegraphics{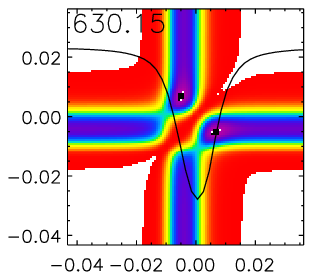}}\hspace*{.75cm}\resizebox{2.5cm}{!}{\includegraphics{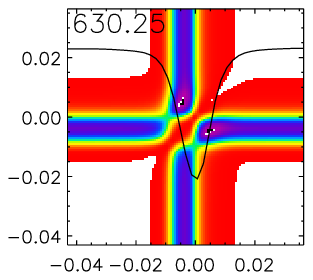}}\hspace*{.2cm}\includegraphics{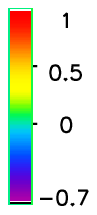}$ $\\$ $\\
\hspace*{.75cm}\resizebox{2.5cm}{!}{\includegraphics{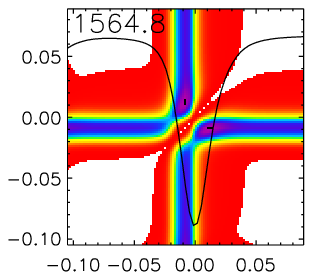}}\hspace*{.75cm}\resizebox{2.5cm}{!}{\includegraphics{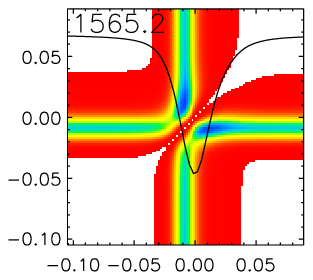}}\hspace*{.2cm}\includegraphics{zbar5.ps}$ $\\
\caption{Correlation matrices from 3-D MHD simulations for 630.15 nm ({\em top left}), 630.25 nm ({\em top right}), 1564.8 nm ({\em bottom left}), and 1565.2 nm ({\em bottom right}).\label{corrmatmhd}}
\end{figure}
\subsection{3-D MHD simulation}
As last numerical model, we obtained spectra in the 1565 nm and 630 nm range
from a 3-D MHD simulation run done with the CO$^5$BOLD code \citep{freytag+etal2002}. The simulations are described in more detail in
\citet{schaffenberger+etal2005,schaffenberger+etal2006}; the simulation box
had an extension of 4.8 Mm x 4.8 Mm in the horizontal domain. The spectra were calculated for a single snapshot of a simulation run that included magnetic
fields; the dynamics in the upper atmosphere is, however, dominated by dynamic
events and shock fronts like in the field-free case \citep[cf.][]{wedemeyer+etal2004}. { The radiative transfer is treaded by mean opacities in the simulation, which prevents to obtain NLTE spectra of, e.g., the chromospheric \ion{Ca}{II} lines directly from the simulation results without additional calculations}; we thus only use the photospheric spectra of \ion{Fe}{I} 630.15 nm, 630.25, 1564.8 nm and 1565.2 nm { that can be synthesized at once}. Figure \ref{corrmatmhd} shows the correlation matrices for these four spectral lines obtained from the 3-D MHD simulation (cp.~to Fig.~\ref{fig6}). The structure of the correlation matrix from the observed spectra is reproduced accurately: the lines with larger line depth (630.15/1564.8 nm) show { a stronger reduction of correlation}, their stripes { with reduced correlation} are broader, and all lines in the matrix from the simulation show the correct red/blue asymmetry. The sole difference to the observations is the { magnitude} of the correlation coefficients: the simulation generally shows stronger anti-correlation (display range from -0.7 instead of -0.3 like for the observations), which presumably is due to the higher spatial resolution of the simulation (40 km grid size). The correlation matrix from the 3-D MHD simulation matches much better to the observations than the simplified 1-D LTE modeling { of wave propagation}, again presumably due to the absence of the granulation pattern in the latter. 
\begin{figure*}
\hspace*{.75cm}\resizebox{12cm}{!}{\includegraphics{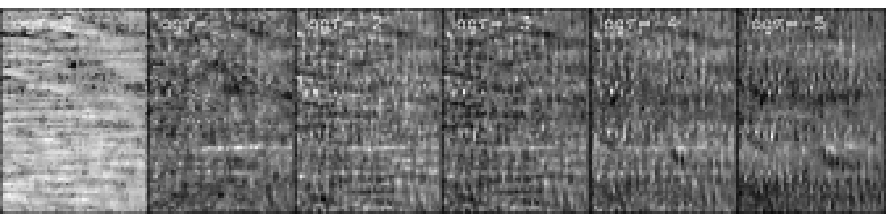}}\hspace*{1.5cm}\resizebox{2.25cm}{!}{\includegraphics{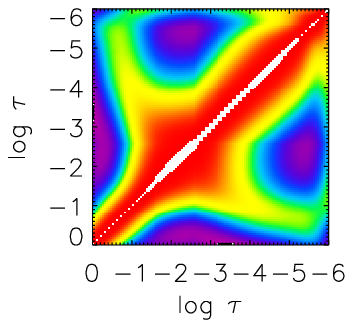}}$ $\\$ $\\$ $\\
\hspace*{.75cm}\resizebox{12cm}{!}{\includegraphics{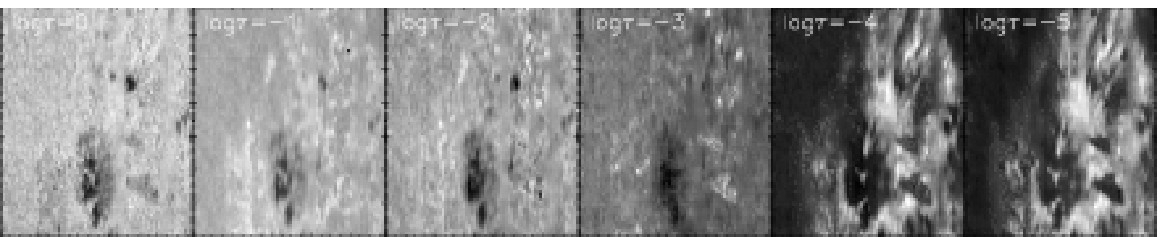}}\hspace*{1.5cm}\resizebox{2.25cm}{!}{\includegraphics{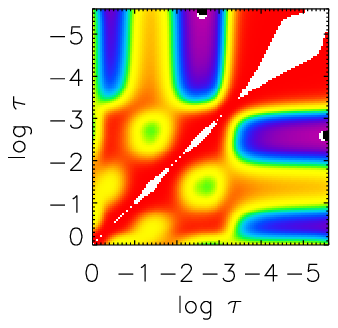}}$ $\\$ $\\$ $\\
\hspace*{.75cm}\resizebox{12cm}{!}{\includegraphics{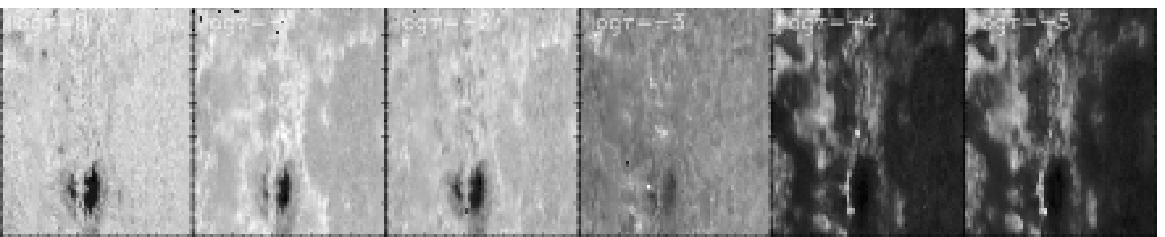}}\hspace*{1.5cm}\resizebox{2.25cm}{!}{\includegraphics{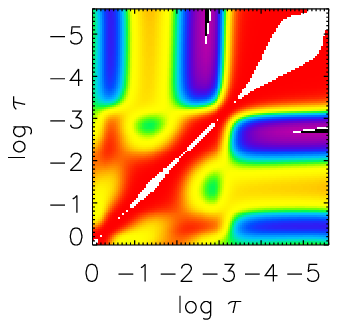}}$ $\\$ $\\$ $\\
\hspace*{.75cm}\resizebox{12cm}{!}{\includegraphics{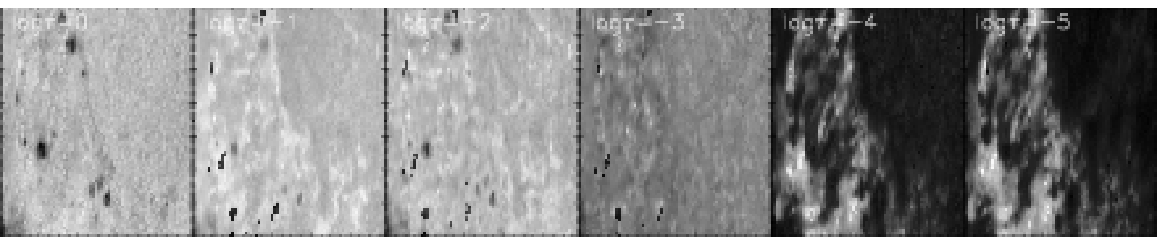}}\hspace*{1.5cm}\resizebox{2.25cm}{!}{\includegraphics{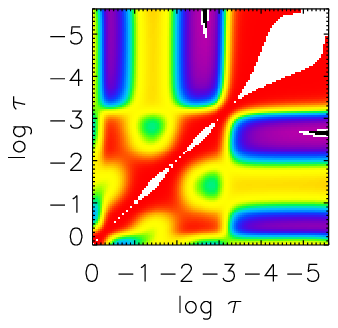}}$ $\\$ $\\$ $\\
\caption{{\em Left top}: temperature maps  at various optical depths from a LTE inversion of \ion{Ca}{II} H spectra of a time-series. {\em Right top}: correlation matrix of temperature as function of optical depth. {\em Bottom three rows}: same for three scans of an active off disc centre.\label{temp_corr}}
\end{figure*} 
\section{Correlation matrices for temperature\label{secttemp}}
To highlight the possible usefulness of the correlation matrix for an analysis
of also other physical quantities than spectra, we also calculated correlation
matrices for temperature as function of optical depth. The temperature values
were derived by an inversion of \ion{Ca}{II} H spectra of an 1-hour
time-series \citep[cf.][observation nr.~21 in the present paper, cf.~Appendix
\ref{obstippolis}]{beck+etal2008} and three scans of an active region off disc
centre (taken on 08/12/2007, $\theta\sim 50^\circ$; not listed in the appendix). The inversion is based on the SIR code in synthesis mode \citep{cobo+toroiniesta1992,cobo1998}; it will be discussed in more detail in a forthcoming paper. It mainly differs from the standard version of SIR in the usage of a pre-calculated {\em fixed} intensity-temperature response function, and the initial use of an archive of pre-calculated Ca spectra. The usage of the archive was found to be necessary to achieve a reasonable match of the Ca line core region; the standard iterative approach of SIR failed there because of the complex temperature stratifications required. The inversion with the LTE assumption basically maps the intensity values at a given wavelength in the Ca profile to some range in optical depth, like given by the intensity contribution function \citep[cf.~Fig.~5 of][{ A\&A, in press}]{rezaei+etal2008,beck+etal2009}.

Figure \ref{temp_corr} shows the resulting temperature maps at six levels
(log$\tau$=0,-1,...,-5) {\em at left}. The temperature maps of the time-series
are to first order identical to the intensities in the spectral windows OW,
MW2, MW3, IW1 H$_{2}$V and Ca line core in \citet[][Fig.~4]{beck+etal2008}. Correlating the maps of $T(\tau_i, x,t)$ and $T(\tau_j, x,t)$ using
Eq.~(\ref{eq1}) yielded the correlation matrices shown at the {\em right} in
Fig.~\ref{temp_corr}. For the time-series, the shape of the temperature
correlation does not match well to any pattern in the \ion{Ca}{II} H matrix,
but the matrix of \ion{Ca}{II} IR at 854 nm (Fig.~\ref{caircorrmat}) shows
some similarity. The structure seen from $\Delta\lambda= -0.05$ nm to 0 nm in the matrix from the spectra matches to that from log $\tau$= -3 to -6 in temperature: an iterative sequence of increased and reduced width of the high correlation band around the diagonal with lower correlation at, e.g., $r_{\Delta\lambda= -0.05, 0}$ and $r_{\log\tau=-3,-5}$, respectively. The matrices from the temperature maps of the active region
 are similar to each other, but differ from the one for quiet Sun, which will be due to the different heliocentric angle and the solar surface structure { like for the two observations of \ion{Ca}{II} IR 854 nm}. They only keep a few common properties, like, e.g., the anti-correlation at around log$\tau$=-2.5 to all regions above. A conversion from log $\tau$ to the
corresponding wavelength (or vice versa) will be needed for a direct comparison
with the correlation matrices from spectra, or a conversion from $T(\tau)$ to geometrical height for a comparison with simulations, but the inversion code first has to be improved for a quantitative analysis. 
\section{Summary and discussion\label{finalsect}}
We have calculated the linear correlation coefficient between intensities at different wavelengths for various photospheric and chromospheric spectral lines from near-UV to near-IR wavelengths. The correlation coefficients yield a matrix that contains information on the causal relationship between different wavelengths, and thus, height layers in the solar atmosphere. For all photospheric spectral lines, a pronounced asymmetry between the red and blue wing is found. The { intensity at the} line core { has a weaker correlation} with the { intensities in the} red wing than with { those in} the blue one. Correlations between { the intensities at} continuum wavelengths are always high. All chromospheric spectral lines considered (\ion{Ca}{II} H,\ion{Ca}{II} IR at 854 nm and 866 nm) show much more structure in the correlation matrix near their respective line cores than the photospheric lines. 

{ A simple granulation simulation that assumes hot upflows and cold downflows produces a strong reduction of the correlation between line-core intensities of photospheric lines and the intensity at continuum wavelengths. The resulting correlation matrix, however, is fully symmetric in red and blue wing.} Synthetic LTE spectra corresponding to the upwards passage of a temperature
perturbation are able to reproduce the { additional} signature seen in the photospheric lines{, the pronounced asymmetry between red and blue wing}, if a positive velocity (downflow) is introduced together with the
temperature enhancement. Spectra of \ion{Ca}{II} H synthesized from 1-D NLTE
simulations, where propagating waves are generated by a photospheric piston,
yield a correlation matrix that qualitatively matches to the pattern in the
correlation matrix of observed spectra near the very line core of \ion{Ca}{II}
H, but only for the case of a monochromatic driver with 45 sec
period. Stochastic excitation of the piston produces a correlation matrix that
is inconsistent with the observations. A 3-D MHD simulation run, in which the
dynamics in the upper atmosphere is dominated by shock waves, faithfully
reproduces the asymmetry of red and blue wing seen for all photospheric lines,
together with the relative amount of correlation that seems to depend on the
line depth of the respective lines. { The 3-D simulations contain contributions from both granulation and waves.} Temperature maps at different optical
depth layers that were derived from an LTE inversion of observed spectra yield
a correlation matrix that has at least a structural resemblance to the
matrices obtained from observed spectra of \ion{Ca}{II} IR 854 nm. A clear difference between quiet Sun and active regions is seen in the temperature correlation.

\paragraph{Influence of granulation pattern} { The granulation pattern leaves its imprint on the observed correlation matrices in all photospheric spectral lines, which partially hides the signature of the dynamic processes relevant for the chromosphere. The observed asymmetry between the intensities of red and blue wing, however, could not be related to the granulation pattern (Sect.~\ref{gransimm}). To better isolate the dynamic processes, it seems advisable to partly remove the contribution of to the granulation pattern from the spectra prior to the calculation of wavelength correlation matrices in future studies. For large-area maps, this could be achieved by a local (or global) adjustment of continuum intensity values. One could force all spectra in a small-scale (1-3$^{\prime\prime}$ radius) surrounding of each pixel to an identical intensity in continuum wavelengths, corresponding to removing the granulation pattern locally, and then calculate several ``local'' correlation matrices to be averaged later on, or force the continuum intensity to unity across the full FOV. Time-series will be even better suited for the removal of the granulation pattern, because the granulation can be more accurately filtered in the temporal than in the spatial domain: the typical time-scale of granules (5 min) differs from the fast evolving and propagating waves \citep[around 1 min typical life time from first visibility in continuum layers to shock signature in the chromosphere, e.g.,][]{beck+etal2008}, whereas the spatial scales of granules and chromospheric brightenings can be of comparable size. The signature of the dynamic processes will then become more prominent in granulation-filtered data. Preliminary tests with, e.g., the method of local intensity balancing significantly reduced the areas of high correlation around the diagonal.
\section{Conclusions\label{conclusions}}
The matrices of the linear correlation coefficient of photospheric and chromospheric spectral lines carry information on the physical processes in the solar atmosphere. They are influenced by both the convective granulation pattern and more transient events like propagating waves that are presumably related to the chromospheric heating process. We have shown that the most pronounced feature in the correlation matrices of photospheric lines, an asymmetry between the intensities in the red and blue wing, can be reproduced to first order} by assuming propagating (acoustic) waves that produce an intensity enhancement together with a red-shift of the spectral lines. { Such propagating waves also qualitatively reproduce the correlation matrix of chromospheric spectral lines.} 

The calculation of correlation matrices can be applied to other physical quantities as well, like for example temperature maps at different height levels in the solar atmosphere. The method seems suited to compare observations of chromospheric spectral lines, where NLTE conditions apply, with theoretical or numerical modeling of wave propagation in the solar atmosphere. The analysis of observed spectra can be refined by filtering out the signature of the granulation pattern that differs mainly in its temporal behavior from the more transient chromospheric heating process. Other physical quantities like 3-D temperature cubes can be analyzed with the correlation method as well, which seems to be an useful tool for a comparison of simulations and observations.
\begin{acknowledgements}
The authors want to thank F.~W{\"o}ger (NSO) and A.~Tritschler (NSO) for the IBIS data set at 854 nm. Similar thanks go to O.~Steiner (KIS) and R.~Rezaei (KIS) for the 3-D MHD spectra. The VTT is operated by the Kiepenheuer-Institut f\"ur Sonnenphysik (KIS) at the Spanish Observatorio del Teide of the Instituto de Astrof\'{\i}sica de Canarias (IAC). The POLIS instrument has been a joint development of the High Altitude Observatory (Boulder, USA) and the KIS. The National Solar Observatory (NSO) is operated by the Association of Universities for Research in Astronomy, Inc.~(AURA), under cooperative agreement with the National Science Foundation. IBIS has been built by INAF/Osservatorio Astrofisico di Arcetri with contributions from the Universities of Firenze and Roma ``Tor Vergata'', the NSO, and the Italian Ministries of Research (MIUR) and Foreign Affairs (MAE). W.R.~acknowledges support by the Deutsche Forschungsgemeinschaft under grant SCHM 1168/6-1.\\
C.B.~(born 1975) wants to apologize sincerely for not having cited Liu's important work before, he was not aware of its existence.
\end{acknowledgements}
\bibliographystyle{aa}
\bibliography{references_luis_mod}

\begin{thebibliography}{26}
\expandafter\ifx\csname natexlab\endcsname\relax\def\natexlab#1{#1}\fi

\bibitem[{{Beck} {et~al.}(2007){Beck}, {Bellot Rubio}, {Schlichenmaier}, \&
  {S{\"u}tterlin}}]{beck+etal2007}
{Beck}, C., {Bellot Rubio}, L.~R., {Schlichenmaier}, R., \& {S{\"u}tterlin}, P.
  2007, \aap, 472, 607

\bibitem[{{Beck} {et~al.}(2009){Beck}, {Khomenko}, {Rezaei}, \&
  {Collados}}]{beck+etal2009}
{Beck}, C., {Khomenko}, E., {Rezaei}, R., \& {Collados}, M. 2009, \aap,
  submitted

\bibitem[{{Beck} \& {Rezaei}(2009)}]{beck+rezaei2009}
{Beck}, C. \& {Rezaei}, R. 2009, \aap, 502, 969

\bibitem[{{Beck} {et~al.}(2005){Beck}, {Schmidt}, {Kentischer}, \&
  {Elmore}}]{beck+etal2005b}
{Beck}, C., {Schmidt}, W., {Kentischer}, T., \& {Elmore}, D. 2005, \aap, 437,
  1159

\bibitem[{{Beck} {et~al.}(2008){Beck}, {Schmidt}, {Rezaei}, \&
  {Rammacher}}]{beck+etal2008}
{Beck}, C., {Schmidt}, W., {Rezaei}, R., \& {Rammacher}, W. 2008, \aap, 479,
  213

\bibitem[{{Cabrera Solana} {et~al.}(2005){Cabrera Solana}, {Bellot Rubio}, \&
  {del Toro Iniesta}}]{cabrera+bellot+iniesta2005}
{Cabrera Solana}, D., {Bellot Rubio}, L.~R., \& {del Toro Iniesta}, J.~C. 2005,
  \aap, 439, 687

\bibitem[{{Carlsson} \& {Stein}(1997)}]{carlsson+stein1997}
{Carlsson}, M. \& {Stein}, R.~F. 1997, \apj, 481, 500

\bibitem[{{Cavallini}(2006)}]{cavallini2006}
{Cavallini}, F. 2006, \solphys, 236, 415

\bibitem[{{Collados} {et~al.}(2007){Collados}, {Lagg}, {D{\'{\i}}az
  Garc{\'{\i}} A}, {Hern{\'a}ndez Su{\'a}rez}, {L{\'o}pez L{\'o}pez}, {P{\'a}ez
  Ma{\~n}{\'a}}, \& {Solanki}}]{collados+etal2007}
{Collados}, M., {Lagg}, A., {D{\'{\i}}az Garc{\'{\i}} A}, J.~J., {et~al.} 2007,
  in Astronomical Society of the Pacific Conference Series, Vol. 368, The
  Physics of Chromospheric Plasmas, ed. P.~{Heinzel}, I.~{Dorotovi{\v c}}, \&
  R.~J. {Rutten}, 611

\bibitem[{{Cram}(1978)}]{cram1978}
{Cram}, L. 1978, \aap, 70, 345

\bibitem[{{Freytag} {et~al.}(2002){Freytag}, {Steffen}, \&
  {Dorch}}]{freytag+etal2002}
{Freytag}, B., {Steffen}, M., \& {Dorch}, B. 2002, Astronomische Nachrichten,
  323, 213

\bibitem[{{Gingerich} {et~al.}(1971){Gingerich}, {Noyes}, {Kalkofen}, \&
  {Cuny}}]{gingerich+etal1971}
{Gingerich}, O., {Noyes}, R.~W., {Kalkofen}, W., \& {Cuny}, Y. 1971, \solphys,
  18, 347

\bibitem[{{Holweger} \& {Mueller}(1974)}]{holweger+mueller1974}
{Holweger}, H. \& {Mueller}, E.~A. 1974, \solphys, 39, 19

\bibitem[{{Huenerth} \& {Ulmschneider}(1995)}]{huenerth+ulmschneider1995}
{Huenerth}, G. \& {Ulmschneider}, P. 1995, \aap, 293, 166

\bibitem[{{Liu}(1974)}]{liu1974}
{Liu}, S.-Y. 1974, \apj, 189, 359

\bibitem[{{Rammacher} \& {Ulmschneider}(1992)}]{rammacher+ulmschneider1992}
{Rammacher}, W. \& {Ulmschneider}, P. 1992, \aap, 253, 586

\bibitem[{{Rammacher} \& {Ulmschneider}(2003)}]{rammacher+ulmschneider2003}
{Rammacher}, W. \& {Ulmschneider}, P. 2003, \apj, 589, 988

\bibitem[{{Rezaei} {et~al.}(2008){Rezaei}, {Bruls}, {Schmidt}, {Beck},
  {Kalkofen}, \& {Schlichenmaier}}]{rezaei+etal2008}
{Rezaei}, R., {Bruls}, J.~H.~M.~J., {Schmidt}, W., {et~al.} 2008, \aap, 484,
  503

\bibitem[{{Ruiz Cobo}(1998)}]{cobo1998}
{Ruiz Cobo}, B. 1998, \apss, 263, 331

\bibitem[{{Ruiz Cobo} \& {del Toro Iniesta}(1992)}]{cobo+toroiniesta1992}
{Ruiz Cobo}, B. \& {del Toro Iniesta}, J.~C. 1992, \apj, 398, 375

\bibitem[{{Rutten} {et~al.}(2004){Rutten}, {de Wijn}, \&
  {S{\"u}tterlin}}]{rutten+etal2004a}
{Rutten}, R.~J., {de Wijn}, A.~G., \& {S{\"u}tterlin}, P. 2004, \aap, 416, 333

\bibitem[{{Rutten} \& {Uitenbroek}(1991)}]{rutten+uitenbroek1991}
{Rutten}, R.~J. \& {Uitenbroek}, H. 1991, \solphys, 134, 15

\bibitem[{{Schaffenberger} {et~al.}(2005){Schaffenberger},
  {Wedemeyer-B{\"o}hm}, {Steiner}, \& {Freytag}}]{schaffenberger+etal2005}
{Schaffenberger}, W., {Wedemeyer-B{\"o}hm}, S., {Steiner}, O., \& {Freytag}, B.
  2005, in ESA Special Publication, Vol. 596, Chromospheric and Coronal
  Magnetic Fields, ed. D.~E. {Innes}, A.~{Lagg}, \& S.~A. {Solanki}

\bibitem[{{Schaffenberger} {et~al.}(2006){Schaffenberger},
  {Wedemeyer-B{\"o}hm}, {Steiner}, \& {Freytag}}]{schaffenberger+etal2006}
{Schaffenberger}, W., {Wedemeyer-B{\"o}hm}, S., {Steiner}, O., \& {Freytag}, B.
  2006, in Astronomical Society of the Pacific Conference Series, Vol. 354,
  Solar MHD Theory and Observations: A High Spatial Resolution Perspective, ed.
  J.~{Leibacher}, R.~F. {Stein}, \& H.~{Uitenbroek}, 345

\bibitem[{{Ulmschneider}(1994)}]{ulmschneider1994}
{Ulmschneider}, P. 1994, \aap, 288, 1021

\bibitem[{{Wedemeyer} {et~al.}(2004){Wedemeyer}, {Freytag}, {Steffen},
  {Ludwig}, \& {Holweger}}]{wedemeyer+etal2004}
{Wedemeyer}, S., {Freytag}, B., {Steffen}, M., {Ludwig}, H.-G., \& {Holweger},
  H. 2004, \aap, 414, 1121

\end{thebibliography}
\clearpage
\begin{appendix}
\section{Used observations\label{appa}}
\begin{figure}
\center
\resizebox{6cm}{!}{\includegraphics{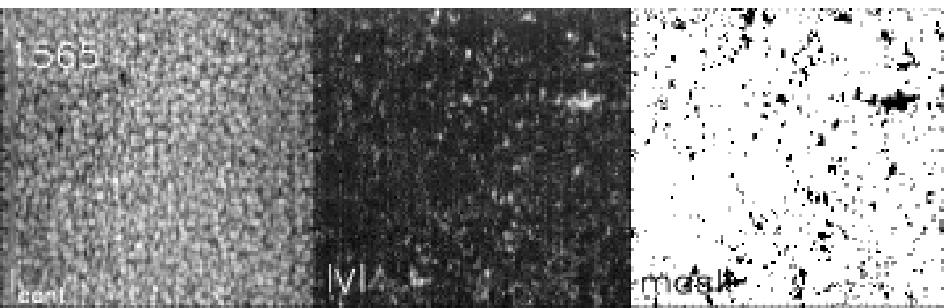}}\\$ $\\
\resizebox{5.6cm}{!}{\includegraphics{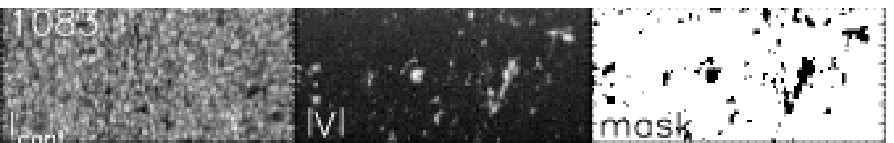}}\\$ $\\
\resizebox{5.6cm}{!}{\includegraphics{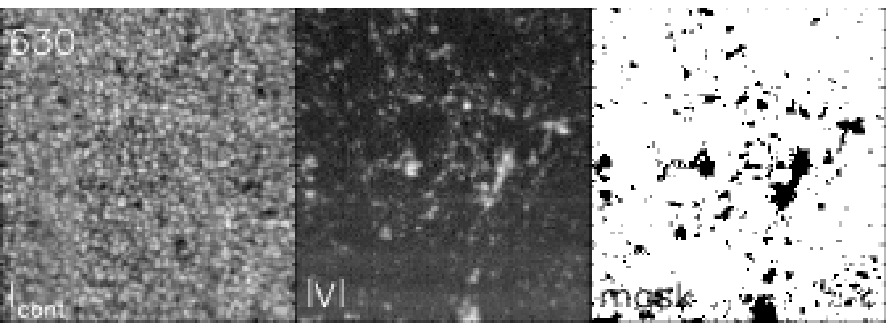}}\\$ $\\
\resizebox{7.5cm}{!}{\includegraphics{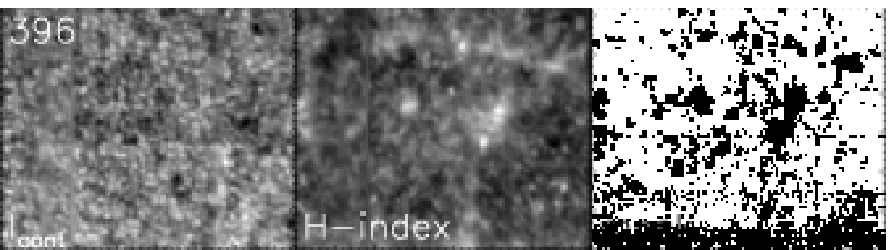}}\\$ $\\
\caption{{ Examples of large-area scans used. {\em Left to right}: continuum
    intensity, absolute integrated Stokes $V$ signal, mask of magnetic
    fields.  {\em Top to bottom}: 1565-6 (see.~Table \ref{tabobs} for the
    details, the number -X denotes the observation); 1083-7; 630-10; 396-15. For \ion{Ca}{II} H, the integrated line core emission (H-index) is shown instead of $V$. \label{figa1}}}
\end{figure}
\begin{figure}
\center
\resizebox{5.5cm}{!}{\includegraphics{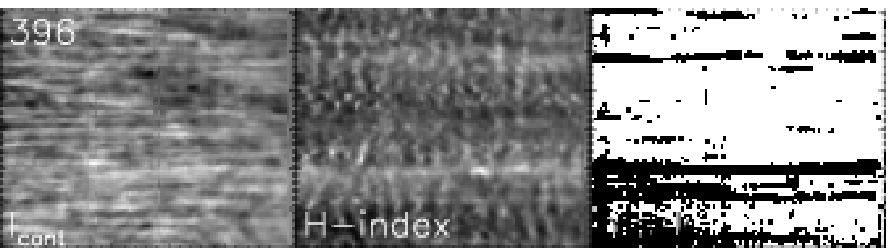}}\\$ $\\
\resizebox{5.5cm}{!}{\includegraphics{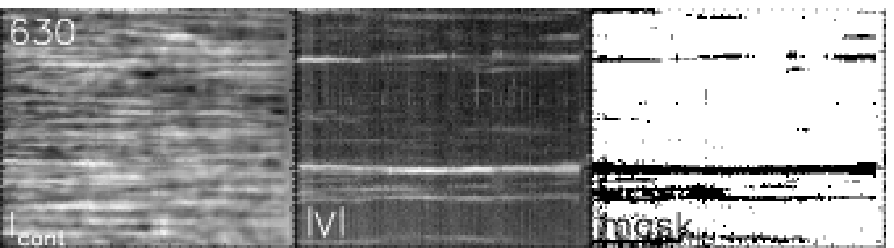}}\\$ $\\
\resizebox{5.5cm}{!}{\includegraphics{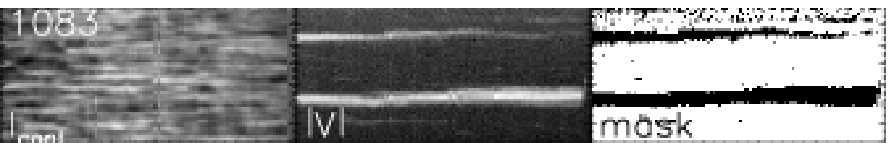}}\\$ $\\
\caption{{ Overview over the time-series used, same layout as Fig.~\ref{figa1}. {\em Top to bottom}: 396-21, 630-22, 1083-23.\label{ts_fig}}}
\end{figure}
\subsection{TIP/POLIS\label{obstippolis}}
For the derivation of the correlation matrices, several data sets taken with
the spectropolarimeters TIP and POLIS were used. Table \ref{tabobs} lists the
data sets with a consecutive number, the operation number of the data set on
the day, the date, the total integration time per slit position, and the
approximate extent of the scanned area. For each of the polarimetric data sets
(TIP at 1083 nm  or 1565 nm, POLIS at 630 nm), we show a continuum intensity
map at left, the absolute wavelength integrated circular polarization in the
middle, and the mask of field-free and magnetic locations derived from the
polarization signal at right. For the TIP data in the 1083 nm range, the
polarization signal of the photospheric \ion{Si}{I} line at 1082.7 nm was used
to define the masks. Some of the TIP observations consist of repeated scans
over the same area, we have not cut them apart for the display but show them
contiguously. For the \ion{Ca}{II} H data from POLIS, the intensity integrated
over 0.1 nm around the Ca line core (H-index) is shown in the middle. The mask
in that case has been created from the simultaneous 630 nm data of the second
POLIS channel. The maps for the time-series used here (Fig.~\ref{ts_fig}) are
organized in the same way as for the large-area scans. Spatial sampling in
scanning direction was 0\farcs5, along the slit the sampling was 0\farcs15
(630 nm), 0\farcs29 (396 nm), 0\farcs17 (1083 and 1565 nm). Additional
information on the respective data sets can be found in their online overview
archives (accessible from http://www.kis.uni-freiburg.de $\rightarrow$
Observatories $\rightarrow$ Data archives).
\begin{table}
\caption{Detailed list of the observations used. The operation number ({\em 2nd column}) is the number of the observation on that day.\label{tabobs}}
\begin{tabular}{llllll}
Nr. & Op. & wavelength & ddmmyy & integ. & size x/y\cr
  & nr & nm &  & time  & in arcsec\cr\hline
1 & 3 & 1565 & 10/08/06 & 30 sec & 35/70  \cr
2 & 3 & 1565 & 12/08/06 & 10 & 120/70 \cr
3 & 1 & 1565 & 14/08/06 & 10 & 200/70 \cr
4 & 1 & 1565 & 21/08/06 & 6 & 75/70 \cr
5 & 1 & 1565 & 07/09/07 & 5 & 80/70 \cr
6 & 7 & 1565 & 07/09/07 & 5 & 80/70 \cr\hline
7 & 5 & 1083 & 24/07/06 & 6.5 & 75/35 \cr
8 & 6 & 1083 & 24/07/06 & 6.5 & 75/35 \cr
9 & 7 & 1083 & 24/07/06 & 6.5 & 75/35 \cr\hline
10 & 6 & 630 & 24/07/06 & 6.5 & 75/87 \cr
11 & 7 & 630 & 24/07/06 & 6.5 & 75/87 \cr
12 & 8 & 630 & 24/07/06 & 6.5 & 75/87 \cr\hline
13 & 2 & 396 & 18/07/06 & 6.5 & 40/63 \cr
14 & 3 & 396 & 18/07/06 & 6.5 & 40/63 \cr
15 & 6 & 396 & 24/07/06 & 6.5 & 75/63 \cr
16 & 7 & 396 & 24/07/06 & 6.5 & 75/63 \cr
17 & 9 & 396 & 08/12/07 & 3.3 & 100/63 \cr\hline
18 & -- & 866 & 20/06/07 & 0.25 & 120/177 \cr\hline
19 & 0 & 854 & 23/09/08 & 100 ms$^1$ & 39/72 \cr
{ 20 }& { - }& { 854} & { 29/08/09} & { 25 sec} &{ 60/84} \cr\hline
21 & 1 & 396 & 24/07/06 & 3.3 & 52$^2$/63 \cr
22 & 1 & 630 & 24/07/06 & 3.3 & 52$^2$/63 \cr
23 & 1 & 1083 & 24/07/06 & 3 & 52$^2$/35 \cr\hline
\end{tabular}\\
\center
$^1$: IBIS data, exposure time per wavelength; $^2$: duration in min
\end{table}
\subsection{Ca II IR\label{obscair}}
\begin{figure}
\centerline{\resizebox{6.cm}{!}{\hspace*{.5cm}\includegraphics{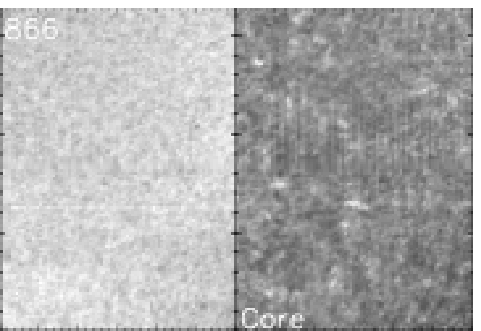}}}$ $\\$ $\\
\centerline{\resizebox{6.cm}{!}{\hspace*{.5cm}\includegraphics{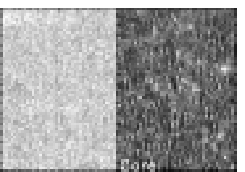}}}$ $\\$ $\\$ $\\
\centerline{\resizebox{8.cm}{!}{\hspace*{.75cm}\includegraphics{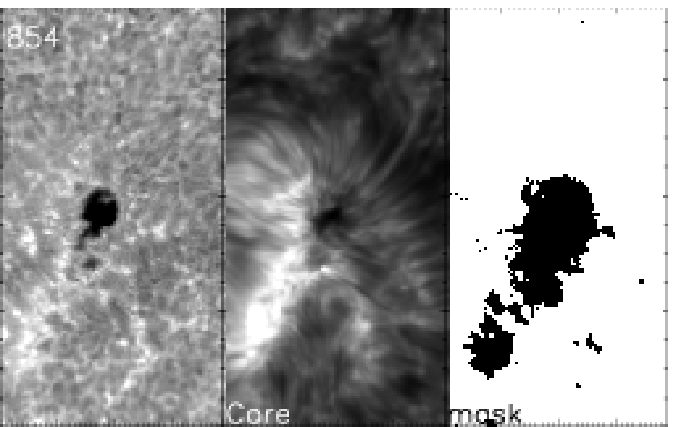}}}$ $\\
\caption{{\em Top}: overview of the \ion{Ca}{II} IR map 866-18. {\em Left}: wing intensity, {\em right}: core intensity. { {\em Middle}: overview of the \ion{Ca}{II} IR map 854-20. {\em Left}: wing intensity, {\em right}: core intensity.} {\em Bottom}: overview of the \ion{Ca}{II} IR map 854-19. {\em Left to right}: wing intensity, core intensity, mask.\label{cairfov}}
\end{figure}

The spectroscopic data of \ion{Ca}{II} IR 866 nm was taken at the VTT using a PCO camera at the main spectrograph. Simultaneously with the \ion{Ca}{II} IR 866 nm line also \ion{Ca}{II} H was observed. A study of centre-to-limb variation and some time series are available from the observation campaign and can be used in further investigations. As no polarimetric data is available in this case, Fig.~\ref{cairfov} only shows a pseudo-continuum map of the intensity in the line wing, and the wavelength integrated line core intensity comparable to the H-index for \ion{Ca}{II} H.

{ For \ion{Ca}{II} IR 854 nm, we used two different data sets. The first map (\ion{Ca}{II} IR 854-20) was obtained again at the VTT using a PCO camera at the main spectrograph. The camera was run in parallel to TIP in \ion{He}{I} 1083 nm and POLIS in its standard configuration; the data were taken in quiet Sun at disc centre. We have not yet aligned the data of all the instruments, so we have not been able to create a mask of locations with significant polarization signal and used the full FOV instead. The second data set for \ion{Ca}{II} IR 854 nm} was taken in September 2008 with the IBIS spectrometer at the Dunn Solar Telescope in Sac Peak/NM. The instrument was used in the spectropolarimetric mode. The mask was in this case derived from the absolute integrated Stokes $V$ signal of \ion{Ca}{II} IR 854 nm for simplicity; a second data set in 630.25 nm taken two minutes later is also available. The 854 nm line was scanned in 30 steps with a spectral sampling of 4.3 pm; spatial sampling was 0\farcs17. The data set was on an active region containing a pore; it is not fully compatible to the quiet Sun observations on disc centre that were used for all other spectral lines. 
\end{appendix}
\end{document}